\documentclass[superscriptaddress,aps,twocolumn,prd,showpacs,nofootinbib]{revtex4-1}
\usepackage{amsmath}
\usepackage{amsfonts}
\usepackage{amssymb}
\usepackage{graphicx}
\makeatletter
\newcommand*{\rom}[1]{\expandafter\@slowromancap\romannumeral #1@}
\makeatother

\usepackage[usenames,dvipsnames]{xcolor}
\usepackage{hyperref}
\hypersetup{colorlinks=true,
citecolor=MidnightBlue, linkcolor=MidnightBlue, urlcolor=MidnightBlue}
\usepackage{tikz}

\newcommand{\be}{\begin{equation}}
\newcommand{\ee}{\end{equation}}
\newcommand{\bea}{\begin{eqnarray}}
\newcommand{\eea}{\end{eqnarray}}

\newcommand{\bseq}{\begin{subequations}}
\newcommand{\eseq}{\end{subequations}}

\allowdisplaybreaks[1]
\newcommand{\orcidicon}{%
	\begin{tikzpicture}
	\draw[lime, fill=lime] (0,0)
		circle [radius=0.16]
		node[white] {{\fontfamily{qag}\selectfont \tiny ID}};
	\draw[white, fill=white] (-0.0625,0.095)
		circle [radius=0.007];
	\end{tikzpicture}	\hspace{-2mm}
}

\newcommand\orcidIsmael{{\href{https://orcid.org/0000-0002-0606-764X}{\orcidicon}}}
\newcommand\orcidFrancisco{{\href{https://orcid.org/0000-0002-9388-8373}{\orcidicon}}}
\newcommand\orcidJosePedro{{\href{https://orcid.org/0000-0002-9758-3366}{\orcidicon}}}

\begin{document}


\title{Wormhole geometries induced by action-dependent Lagrangian theories}

\author{Ismael Ayuso\orcidIsmael\!\!}
\email{ismaelayuso12@gmail.com}
\affiliation{Instituto de Astrof\'{\i}sica e Ci\^{e}ncias do Espa\c{c}o, Faculdade de
Ci\^encias da Universidade de Lisboa, Edif\'{\i}cio C8, Campo Grande,
P-1749-016 Lisbon, Portugal}

\author{Francisco S. N. Lobo\orcidFrancisco\!\!}
\email{fslobo@fc.ul.pt}
\affiliation{Instituto de Astrof\'{\i}sica e Ci\^{e}ncias do Espa\c{c}o, Faculdade de
Ci\^encias da Universidade de Lisboa, Edif\'{\i}cio C8, Campo Grande,
P-1749-016 Lisbon, Portugal}

\author{Jos\'{e} P. Mimoso\orcidJosePedro\!\!}
\email{jpmimoso@fc.ul.pt}
\affiliation{Instituto de Astrof\'{\i}sica e Ci\^{e}ncias do Espa\c{c}o, Faculdade de
Ci\^encias da Universidade de Lisboa, Edif\'{\i}cio C8, Campo Grande,
P-1749-016 Lisbon, Portugal}

\date{\today}

\begin{abstract}

In this work, we explore wormhole geometries in a recently proposed modified gravity theory arising from a non-conservative gravitational theory, tentatively denoted action-dependent Lagrangian theories. The generalized gravitational field equation essentially depends on a background four-vector $\lambda^\mu$, that plays the role of a coupling parameter associated with the dependence of the gravitational Lagrangian upon the action, and may generically depend on the spacetime coordinates. Considering wormhole configurations, by using ``Buchdahl coordinates'', we find that the four-vector is given by $\lambda_{\mu}=\left(0,0,\lambda_{\theta},0\right)$, and that the spacetime geometry is severely restricted by the condition $g_{tt}g_{uu}=-1$, where $u$ is the radial coordinate. We find a plethora of specific asymptotically flat, symmetric and asymmetric, solutions with power law choices for the function $\lambda$, by generalizing the Ellis-Bronnikov solutions and the recently proposed black bounce geometries, amongst others. We show that these compact objects possess a far richer geometrical structure than their general relativistic counterparts. 

\end{abstract}

\maketitle

\section{Introduction}

A key ingredient in traversable wormhole geometries is the flaring-out condition \cite{Morris:1988cz,Morris:1988tu}, which in General Relativity (GR) entails the violation of the null energy condition (NEC). The latter is defined as $T_{\mu\nu}k^{\mu}k^{\nu} \geq 0$, for {\it any} null vector $k^\mu$ \cite{Visser:1995cc,Lobo:2017oab}, and matter violating the NEC has been denoted as {\it exotic matter}. However, it has been shown that these violations may be minimized using several procedures, such as the cut-and-paste techniques in the thin-shell formalism, where the exotic matter is concentrated at the junction interface \cite{Visser:surgical, Visser:examples, Poisson,Lobo:2003xd,Lobo:2004rp,Lobo:2005zu,Montelongo-Garcia:2011, Bouhmadi-Lopez:2014gza, Lobo:2020kxn,Berry:2020tky}. 
In fact, the problem is improved with evolving traversable wormholes, where it has been demonstrated that these time-dependent geometries may satisfy the energy conditions in arbitrary finite 
intervals of time  \cite{Kar:1994tz,Kar:1995ss}, and recently specific dynamical four-dimensional solutions were presented that satisfy the null and weak energy conditions everywhere and everywhen \cite{KordZangeneh:2020jio,KordZangeneh:2020ixt}.
In fact, modified theories of gravity is an interesting avenue of research to explore traversable wormholes, where these compact objects possess a richer geometrical structure than their general relativistic counterparts. In this context, it has been shown that the NEC can be satisfied for normal matter threading the wormhole throat, where it is the higher order curvature terms that sustain the wormhole \cite{modgravity1,modgravity2,modgravity3,modgravity4,Bohmer:2011,Harko:2013,Mehdizadeh:2015jra,Zangeneh:2015jda,Rosa:2018jwp}.

In this work, we will be interested in studying wormhole geometries in a recently proposed modified gravity theory arising from a non-conservative gravitational theory, tentatively denoted action-dependent Lagrangian theories \cite{Lazo:2017udy}. The latter are obtained through an action principle for action-dependent Lagrangians by generalizing the Herglotz variational  problem \cite{Herglotz1,Herglotz2} for several independent variables. The novel feature when comparing with previous implementations of dissipative effects in gravity is the possible arising of such phenomena from a least action principle, so they are of a purely geometric nature. Applications to this model have also been explored, namely, in cosmology \cite{Fabris:2017msx}, braneworld gravity \cite{Fabris:2018nli}, cosmic string configurations \cite{Braganca:2018elt}, the late-time cosmic accelerated expansion and large scale structure \cite{Carames:2018atv}, and static spherically symmetric stellar solutions \cite{Fabris:2019qvy}, amongst others.


The complete set of field equations considered in this action-dependent Lagrangian theory \cite{Lazo:2017udy} is based on the following total Lagrangian
\begin{equation}
 \mathcal{L}=\mathcal{L}_g+\mathcal{L}_m=(R-\lambda_\mu s^\mu)+\mathcal{L}_m,
 \label{Lagrang}
\end{equation}
where the Einstein-Hilbert Lagrangian is extended with the geometrical sector dealing with the additional dissipative term $\lambda_\mu s^\mu$ while $\mathcal{L}_m$ is the Lagrangian of
the matter fields. In general, the background four-vector $\lambda^\mu$ depends on the spacetime coordinates, however it can be assumed to be constant. The field $s^\mu$ is an action-density field which 
disappears after the variation of the action such that the modification to the GR counterpart is given by the four-vector $\lambda^\mu$ only. Note that $\lambda^\mu$ may be considered a background four-vector, that plays the role of a coupling parameter associated with the dependence of the gravitational Lagrangian upon the action. In the majority of the works considered above it is assumed to be constant, however, in a more general scenario, one may assume it to be a coordinate-dependent four-vector.

Thus, the field equations are given by
\begin{equation}\label{fieldeq}
 G_{\mu\nu}+Z_{\mu\nu}=\kappa^2 T_{\mu\nu},
\end{equation}
where we have defined $\kappa^2=8\pi$, $G_{\mu\nu}$ is the Einstein tensor, and for notational simplicity, we have defined $Z_{\mu\nu}$ as 
\begin{equation}
Z_{\mu\nu}=K_{\mu\nu}-\frac{1}{2}g_{\mu\nu}K.
\label{Zdefined}
\end{equation}
The symmetric geometric structure $K_{\mu\nu}$ is defined as
\begin{equation}
 K_{\mu\nu}=\lambda_\alpha\Gamma^\alpha_{\mu\nu}-\frac{1}{2}(\lambda_\mu\Gamma^\alpha_{\nu\alpha}+\lambda_\nu\Gamma^\alpha_{\mu\alpha}),
\end{equation}
which is constructed from the particular combination of the four-vector $\lambda_{\mu}$ and the Christofell symbols 
\begin{equation}
\Gamma^\alpha_{\mu\nu}= \frac{g^{\alpha\beta}}{2}\left(g_{\beta\mu,\nu}+g_{\beta\nu,\mu}-g_{\mu\nu,\beta}\right).
\end{equation} 
The quantity $K_{\mu\nu}$ (and its trace $K$) represents the geometric structure behind the dissipative nature of the theory. Note that the limit of a vanishing $\lambda_{\mu}$ restores the dissipationless feature of GR.

Thus, motivated by the existence of static spherically-symmetric compact objects analysed in \cite{Fabris:2019qvy}, we extend this analysis to the context of wormhole physics. 
This work is outlined in the following manner: In Sec. \ref{S:wormholethroat}, we present the most general restrictions on static and spherically symmetric wormhole geometries imposed by the geometrical structure of the action-dependent Lagrangian theory. In Sec. \ref{Sec:whsolutions}, we consider a plethora of specific solutions of action-dependent Lagrangian induced wormhole geometries. Finally, in Sec. \ref{sec:conclusions}, we summarize our results and conclude.

\section{General restrictions on wormhole geometries}\label{S:wormholethroat}


Consider the general static and spherically symmetric metric given by 
\begin{equation} \label{genmetric}
ds^2=-f(u)dt^2+g(u)du^2+R^2(u)d\Omega^2,
\end{equation}
where $d\Omega^2=d\theta^2+\sin^2\theta d\phi^2$ is the linear element of the unit sphere, and the metric functions $f(u)$, $g(u)$ and  $R(u)$ are functions of the radial coordinate $u$.
The coordinate choices used in metric (\ref{genmetric}) are often called ``Buchdahl coordinates'' \cite{finch-skea,petarpa1,petarpa2}.
Note that one possesses a freedom in choosing the radial coordinate, consequently allowing one to fix the form of one of the metric functions $f(u)$, $g(u)$ or $R(u)$, which will be considered below.
Here the radial coordinate lies in the range $u \in (-\infty, + \infty)$, so that two asymptotically flat regions exist, i.e., $u \rightarrow \pm \infty$, and are connected by the throat. 
The function $R(u)$ possesses a global positive minimum at the wormhole throat $u=u_0$, which one can set at $u_0=0$, without a loss of generality. Thus, the wormhole throat is defined as $R_0={\rm min}\{R(u)\}=R(0)$. 

In order to avoid event horizons and singularities throughout the spacetime, one imposes that the metric functions $f(u)$ and $g(u)$ are positive and regular everywhere. 
Taking into account these restrictions, namely, the necessary conditions for the minimum of the function imposes the flaring-out conditions, which are given by 
\be
R_0'=0\,, \qquad R_0'' > 0\,.  
\label{flaringout}
\ee

In this work, we consider an anisotropic distribution of matter threading the wormhole described by the following stress-energy tensor $T_{\mu\nu}$
\begin{equation}
T_{\mu\nu}=(\rho+p_t)U_\mu \, U_\nu+p_t\,
g_{\mu\nu}+(p_r-p_t)\chi_\mu \chi_\nu \,,
\label{SETprofile}
\end{equation}
where $U^\mu$ is the four-velocity, $\chi^\mu$ is the unit
spacelike vector in the radial direction, i.e., $\chi^\mu=g^{-1/2}(u) \delta^{\mu}_{u}$; $\rho(u)$ is the energy
density, $p_r(u)$ is the radial pressure measured in the direction
of $\chi^\mu$, and $p_t(u)$ is the transverse pressure measured in
the orthogonal direction to $\chi^\mu$.

Now, rather than write out the full gravitational field equations (\ref{fieldeq}) for the metric (\ref{genmetric}), we note that the only non-zero components of the Einstein and the stress-energy tensor are the diagonal terms, so that the non-diagonal part of the additional tensor $Z_{\mu\nu}$, defined by Eq. (\ref{Zdefined}), also provides additional information on the geometrical structure of the solutions of the theory, namely, that $Z_{\mu\nu}=0$ for $\mu\neq\nu$. More specifically, the non-diagonal components of the symmetric tensor $Z_{\mu\nu}$ places restrictions on the form of the four-vector $\lambda_{\mu}$.

The independent components of the tensor $Z^{\mu}_{\nu}=K^{\mu}_{\nu}-\frac{1}{2}\delta^{\mu}_{\nu}K$ are given by
\begin{eqnarray}
Z^{u}_{t} &=& -\frac{\lambda_{t}[fg'R-g(f'R-4fR')]}{4fg^2 R}, \quad
Z^{u}_{u}=\frac{\lambda_{\theta}}{R^2}\cot \theta, 
	\\
Z^{u}_{\theta} &=& -\frac{\lambda_{\theta}(fg)'}{4fg^2} -\frac{\lambda_{u}}{2g}\cot \theta  , \qquad
Z^{u}_{\phi}=-\frac{\lambda_{\phi}(fg)'}{4fg^2}  , \\
Z^{\theta}_{\theta}&=&Z^{\phi}_{\phi} = \frac{\lambda_{u}(Rf '+2fR')}{2fgR}, \\
Z^{\theta}_{\phi}&=&\frac{\lambda_{\phi}}{2R^2}\cot \theta, \qquad  Z^{\theta}_{t}=-\frac{\lambda_{t}}{2R^2}\cot \theta , \\
Z^{t}_{t}&=& \frac{2\lambda_{u}R'}{gR} + \frac{\lambda_{\theta}}{R^2}\cot\theta ,  \qquad Z^{t}_{\phi}=0 \,.
\end{eqnarray}

From $Z^{\theta}_{\phi}=Z^{\theta}_{t}=0$, one readily extracts the restrictions $\lambda_{\phi}=\lambda_{t}=0$. Taking into account the assumption of the static and spherical symmetric character of the spacetime, the field equations should only depend on the radial coordinate, so that from the diagonal $Z^{u}_{u}$ component, one readily verifies that $\lambda_{\theta} \propto (\cot \theta)^{-1} $, or more specifically $\lambda_{\theta} = \lambda(u)/ (\cot \theta)$ (one may consider the simple case $\lambda(u)=\lambda_0={\rm const}$). 
We emphasize that this result was also obtained in \cite{Fabris:2019qvy}.
Note that if one were to consider $\lambda_{\theta}=0$, then the condition $Z^{u}_{\theta}=0$ would impose that $\lambda_u=0$, taking us trivially back to GR. Analogously, in order for $Z^{t}_{t}$ to only depend on the radial coordinate, from $Z^{u}_{\theta}=0$ this imposes that $\lambda_u=0$ and consequently places a further constraint on the metric functions, namely, $(fg)'=0$. 

Thus, the additional information on the geometrical structure of the theory, which imposes that the non-diagonal components of the symmetric tensor $Z_{\mu\nu}$ vanish, imposes the following condition on the four-vector $\lambda_{\mu}$:
\begin{equation}
\lambda_{\mu}=\left(0,0,\frac{\lambda(u)}{ \cot \theta},0\right),
\label{constraintlambda}
\end{equation}
and the additional geometric tensor $Z^{\mu}_{\nu}$ takes the diagonal form $Z^{\mu}_{\nu}=(\lambda(u)/R^2)\,{\rm diag}(1,1,0,0)$. Furthermore, from the constraint on the metric functions $(fg)'=0$, we can consider, without a loss of generality, the following choice:
\begin{equation}
g(u)=f^{-1}(u)=A(u).
\label{constraintmetric}
\end{equation}

\section{Specific solutions of action-dependent Lagrangian induced wormhole geometries}\label{Sec:whsolutions}

The analysis outlined in the previous section imposes that the static and spherical symmetric configuration (\ref{genmetric}) in the theory (\ref{Lagrang}), can be written as 
\begin{equation} \label{metric}
ds^2=-A(u)dt^2+A^{-1}(u)du^2+R^2(u)d\Omega^2,
\end{equation}
where as before the wormhole throat is defined as $R_0={\rm min}\{R(u)\}=R(0)$, and in order to avoid event horizons and singularities throughout the spacetime, one imposes that the function $A(u)$ is positive and regular everywhere. These restrictions imposes the flaring-out conditions, translated by Eq. (\ref{flaringout}).

As the metric function $A(u)$ is positive and regular for $\forall u$, it is useful to analyse its derivatives at the throat $u=0$. In particular, the sign of $A_0''$ determines the type of extrema of $A(u)$, i.e., it is a minimum if $A_0''>0$ and a maximum if $A_0''<0$.
This implies that the maximum (minimum) of $A(u)$ corresponds to a maximum (minimum) of the gravitational potential, so that in the vicinity of a maximum (minimum) the gravitational force is repulsive (attractive). Thus, the wormhole throat possesses a repulsive or an attractive nature that depends on the sign of $A_0''$.

Now, taking into account the modified Einstein equation (\ref{fieldeq}), the spacetime metric (\ref{metric}) and the stress-energy tensor (\ref{SETprofile}), the gravitational field equations are finally given by:
\begin{eqnarray}
8\pi \rho &=& - \frac{2ARR''+ AR'^2 + A' RR' -1}{R^2}- \frac{\lambda(u)}{R^2} \,, 
	\label{rhou} \\
8\pi p_r &=& \frac{AR'^2 + A' RR' -1}{R^2}+ \frac{\lambda(u)}{R^2} \,,
	\label{pru} \\
8\pi p_t &=& \frac{A''R + 2AR'' +2A'R'}{2R} 
	\label{ptu} \,.
\end{eqnarray}

Adding Eqs. (\ref{rhou}) and (\ref{pru}), yields the following relation
\begin{equation}
R''|_{R_0}=-\frac{4\pi R}{A}(\rho + p_r)|_{R_0}\,,
\end{equation}
and using the condition at the throat $R_0''>0$, one verifies that in these specific action-dependent Lagrangian theories the NEC is generically violated at the throat, i.e., $(\rho + p_r)|_{R_0}<0$.

Taking into account the field equations (\ref{rhou})--(\ref{ptu}), one has three independent equations with six unknown functions of the radial coordinate $u$, namely, $\rho(u)$, $p_r(u)$, $p_t(u)$, $A(u)$, $R(u)$ and $\lambda(u)$. There are several strategies that one may now follow. More specifically, one may consider specific choices for the components of the stress-energy tensor, and then solve the field equations to determine the metric functions and $\lambda(u)$; one may also take into account a plausible stress-energy tensor profile by imposing equations of state $p_r=p_r(\rho)$ and $p_t=p_t(\rho)$, and close the system by adequately choosing the energy density, or any of the metric functions. In alternative to this approach, one may use the reverse philosophy usually adopted in wormhole physics by simple choosing specific choices for the metric functions and $\lambda(u)$, and through the field equations determine the stress-energy profile responsible for sustaining the wormhole geometry. In the following section, we will adopt several strategies outlined above, and a mixture thereof, to obtain specific exact solutions of wormhole spacetimes induced by these action-dependent Lagrangian theories.

\subsection{Specific wormhole solutions: Ellis-Bronnikov solution}\label{elliswh}

\begin{figure*}[htbp!]
    \centering
    \includegraphics[width=0.47\linewidth]{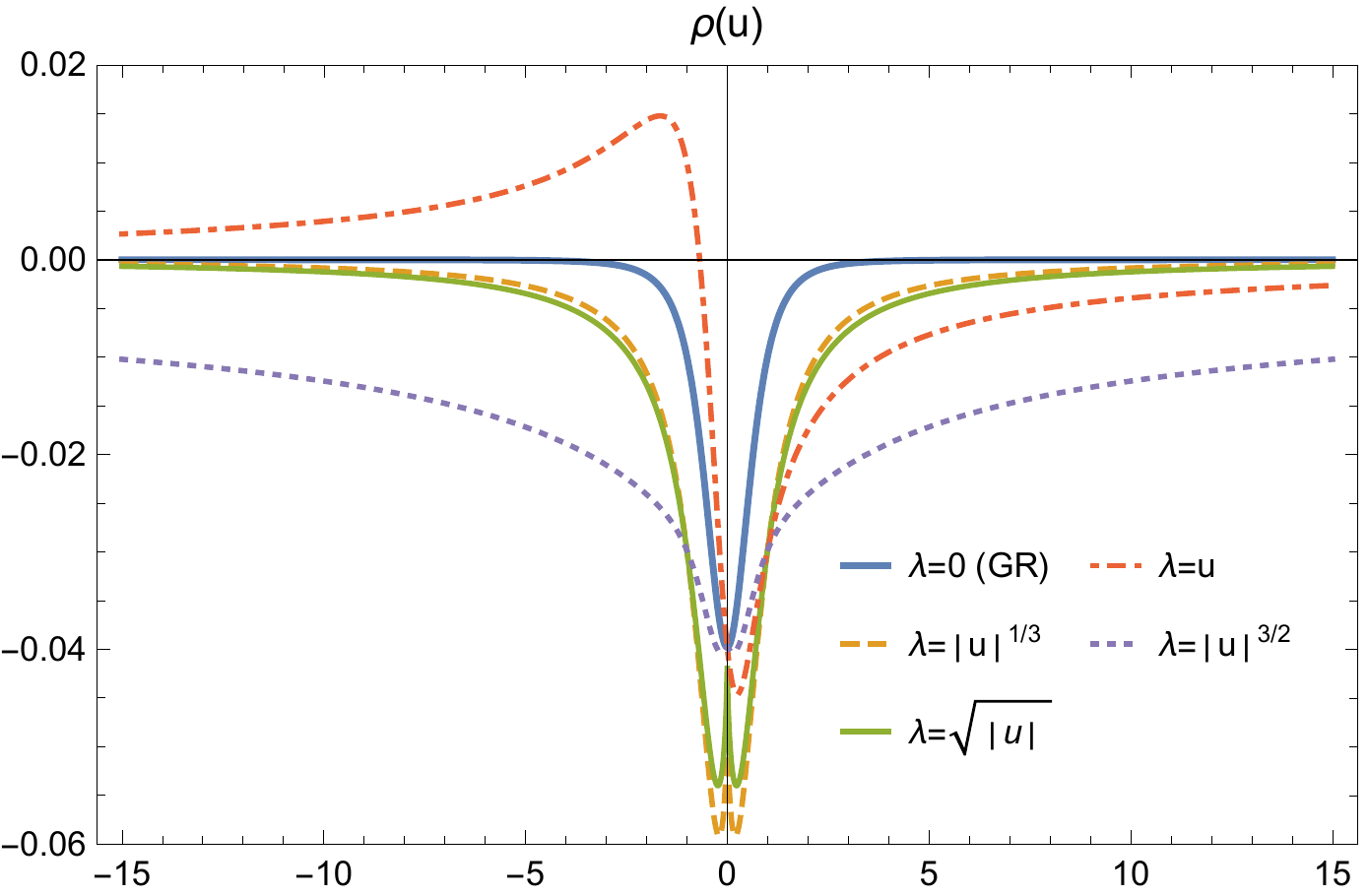}
    \includegraphics[width=0.47\linewidth]{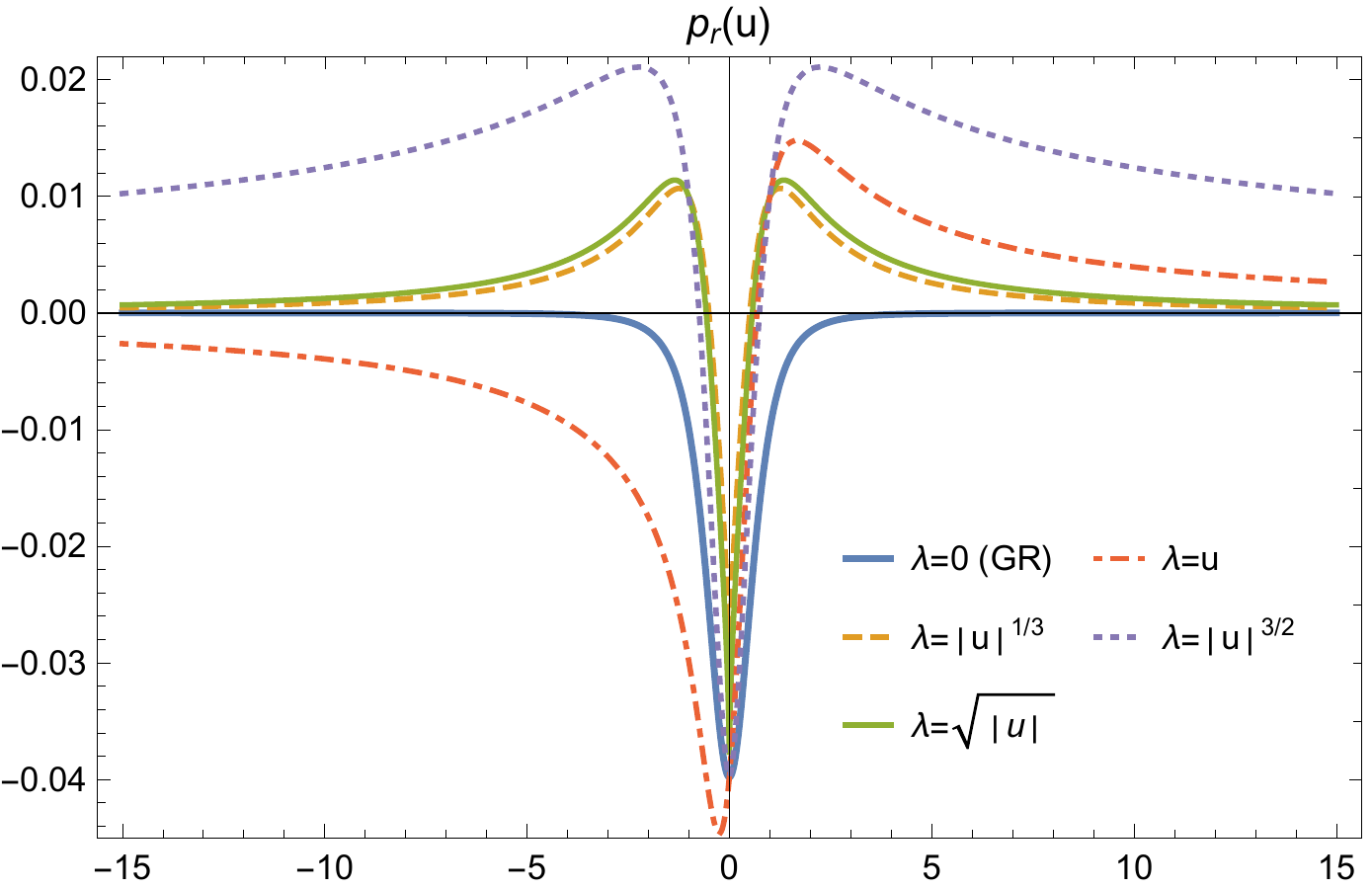}
    \includegraphics[width=0.47\linewidth]{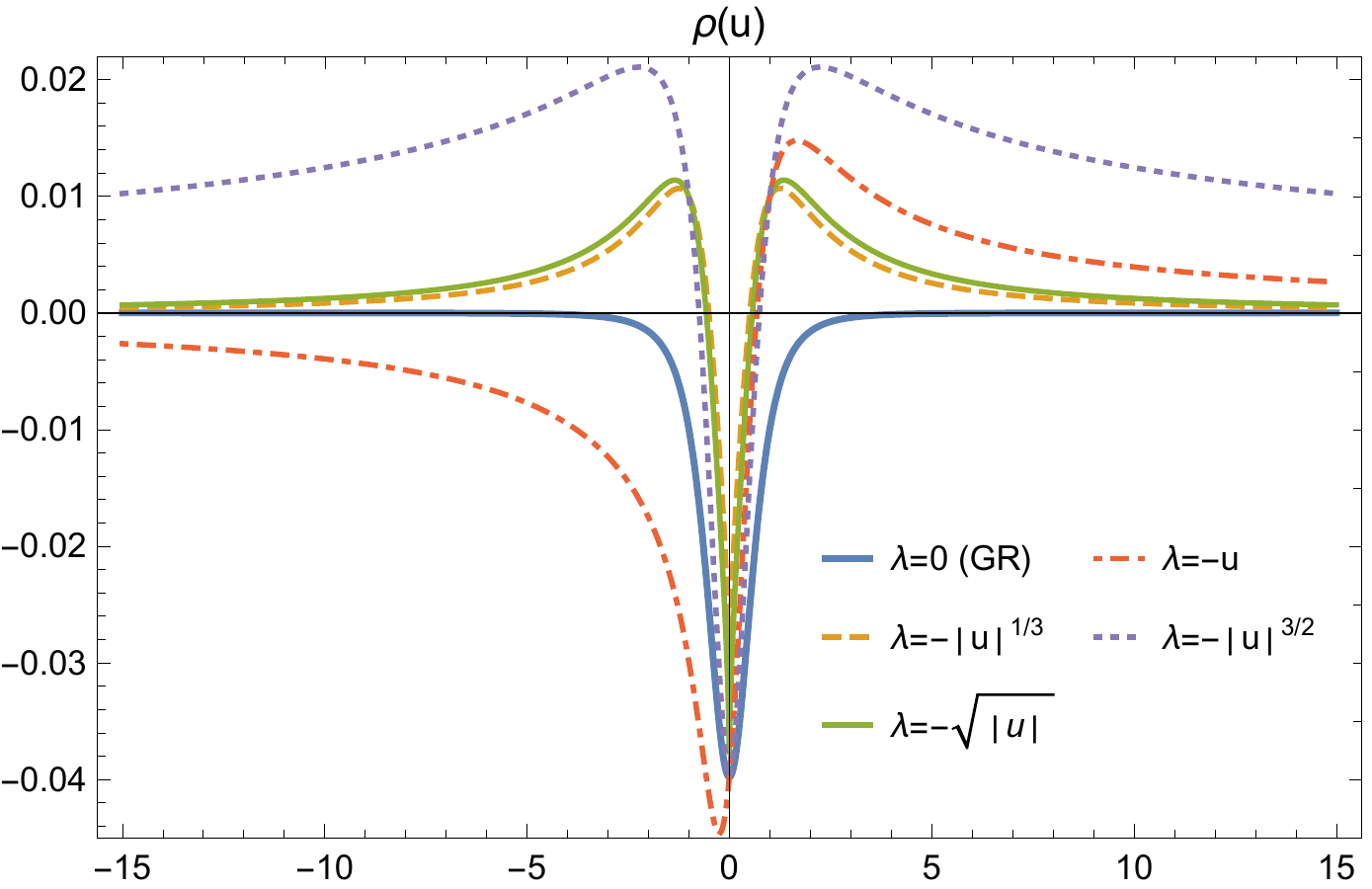}
    \includegraphics[width=0.47\linewidth]{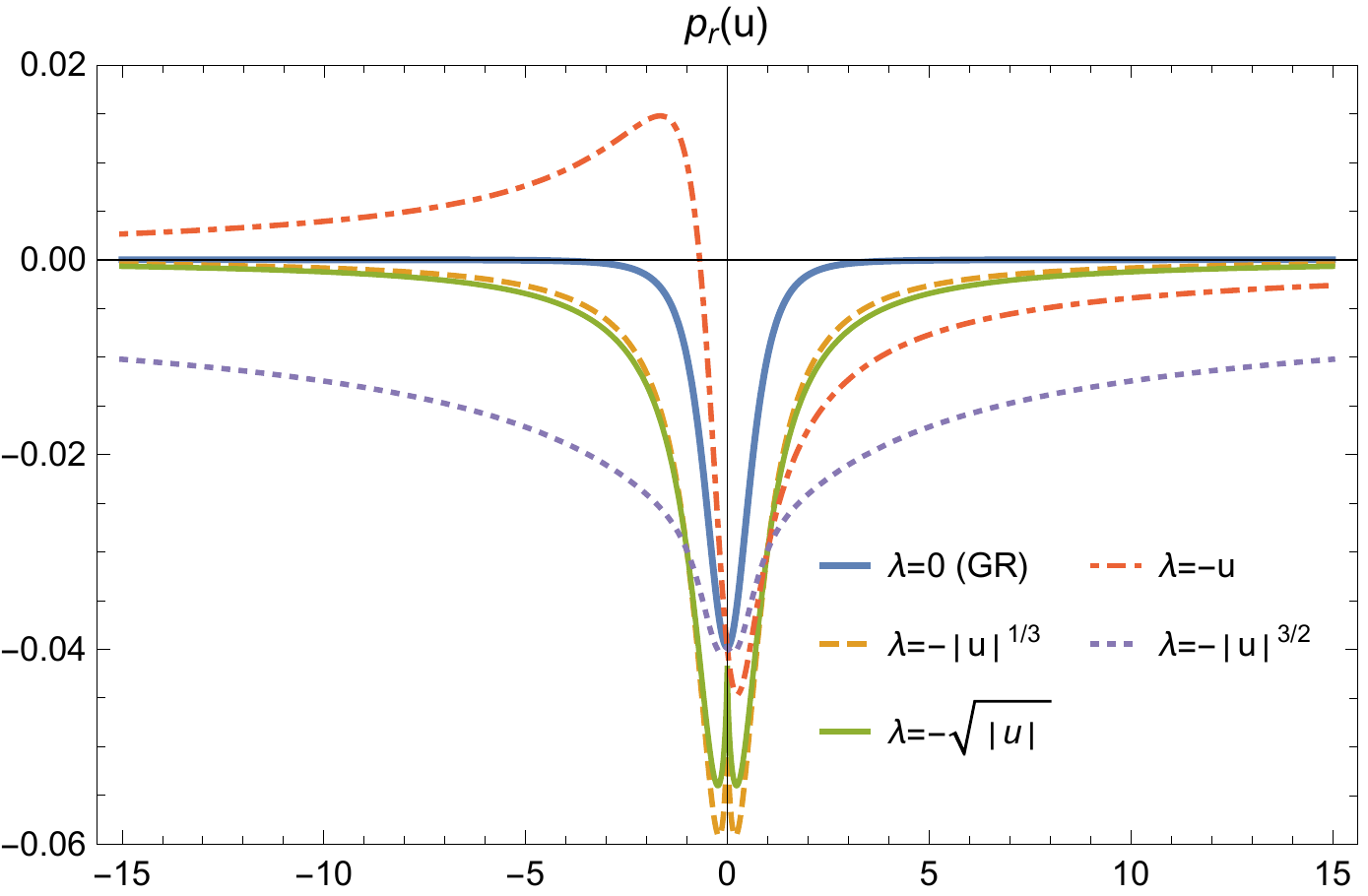}
    \caption{The plots depict the specific case of the Ellis-Bronnikov wormhole configuration with $a=1$, for convenience, and for different choices of the function $\lambda(u)$. 
Depending on the sign of $\lambda(u)$, one obtains a plethora of specific symmetric or asymmetric solutions. Note that for $\lambda(u) \sim \pm u$, one obtains asymmetric solutions, where the energy density is negative at the throat and in the positive (negative) branch of $u$, but becomes positive in the negative (positive) branch; the radial pressure exhibits the inverse qualitative behavior. We refer the reader to the text for more details.}
    \label{GraphEB}
\end{figure*}

Using an appropriate parametrization, we can present a solution by taking into account the reverse philosophy of solving the modified field equations, as follows \cite{Sushkov:2011jh}
\be\label{KGB_metric_func}
R(u)=e^{-\alpha(u)}\sqrt{u^2+a^2}, \qquad A(u)=e^{2\alpha(u)},
\ee
with the factor $\alpha(u)$ defined as
\be 
\alpha(u)=\left(\frac{m}{a}\right) \arctan\left(\frac{u}{a}\right),
\ee
where $m$ and $a$ are two free parameters. Thus, the spacetime metric is given by
\be
ds^2=-e^{2\alpha}dt^2+e^{-2\alpha}\left[du^2 +(u^2+a^2)d\Omega^2 \right].
\ee   

Following the previous definition of the wormhole throat, which is situated at $u_0=0$,
we readily obtain $R'(0)=-m/a$, so that the condition $R'(0)=0$ imposes $m=0$.  Note that these conditions imply that the solution reduces to the well-known Ellis-Bronnikov wormhole spacetime \cite{homerellis,homerellis2,bronikovWH}. This does indeed simplify the analysis below, such that $\alpha(u)=0$, $A(u)=1$, and
\be
R''(u_0)=\frac{1}{|a|}>0
\ee

In addition to this, Eqs. \eqref{rhou}-\eqref{ptu} yield the following stress-energy profile:
\begin{eqnarray}
\rho(u)&=&-\frac{\left(a^2+u^2\right) \lambda (u)+a^2}{8 \pi  \left(a^2+u^2\right)^2},
	\label{EBrho}	\\
p_r(u)&=&\frac{\left(a^2+u^2\right) \lambda (u)-a^2}{8 \pi  \left(a^2+u^2\right)^2},
	\label{EBpr} \\
p_t(u)&=&\frac{a^2}{8 \pi  \left(a^2+u^2\right)^2} \,.
\end{eqnarray}
For the specific case of $\lambda=0$, where the four-vector $\lambda^\mu$ vanishes, this solution simply reduces to the general relativistic Ellis-Bronnikov stress-energy components. However, for the general case, one still needs to impose one more condition to close the system, and in the following we consider specific choices for the function $\lambda(u)$. Equations (\ref{EBrho})--(\ref{EBpr}) yield the following relation:
\be
\rho(u)+p_r(u)=-\frac{a^2}{4 \pi  \left(a^2+u^2\right)^2}\,,
\ee
which states that the NEC is violated throughout the entire spacetime, and is independent of the function $\lambda(u)$.

We are only interested in asymptotically flat solutions, so that taking into account the limit of Eq. (\ref{EBrho}), one finds 
\be
\lim_{u\rightarrow\infty} \rho(u) \sim - \, \lim_{u\rightarrow\infty}  \frac{ \lambda(u) }{u^2} \,.
\ee
For instance, assuming a power law solution for $\lambda(u) \sim u^\alpha$, the asymptotic flatness condition imposes that $\alpha <2$, and from the regularity of the stress-energy components we have $\alpha\geq 0$, so that the parameter lies in the range $0 \leq \alpha <2$. One may perform a similar analysis with the radial pressure $p_r(u)$, but with a change in the sign for the limit. Note that the tangential pressure, $p_t(u)$, is independent of $\lambda(u)$, possesses a maximum value at the throat, $p_t(u=0)= (8 \pi  a^2)^{-1}$, and tends to zero with increasing $u$.

Several choices for the function $\lambda(u)$ are depicted in Fig. \ref{GraphEB}. Depending on the sign of $\lambda(u)$, one obtains a plethora of specific symmetric or asymmetric solutions. More specifically, for the case of $\lambda(u) \sim \pm u$, one obtains asymmetric solutions where the energy density is negative at the throat and in the positive (negative) branch of $u$, but becomes positive in the negative (positive) branch, while the radial pressure possesses the inverse qualitative behavior, as is transparent from Fig. \ref{GraphEB}. Thus, it is possible to alleviate the negative energy densities needed to thread this wormhole configurations, relative to GR. The wormhole solutions obtained with $\lambda \neq 0$ possess a richer structure that their general relativistic counterparts.

\subsection{Specific stress-energy profile}

\begin{figure*}[htbp!]
    \centering
    \includegraphics[width=0.47\linewidth]{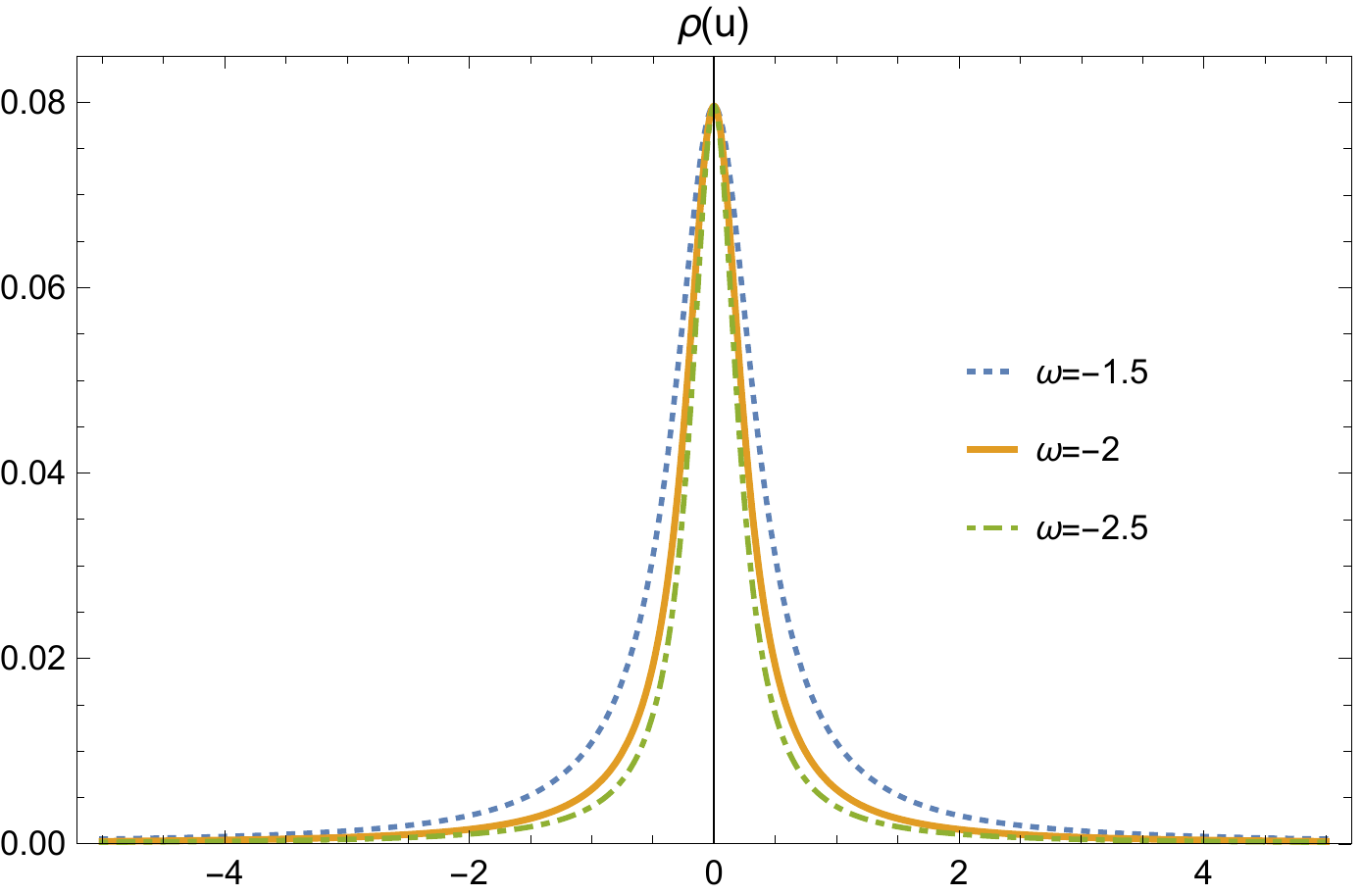}
    \includegraphics[width=0.47\linewidth]{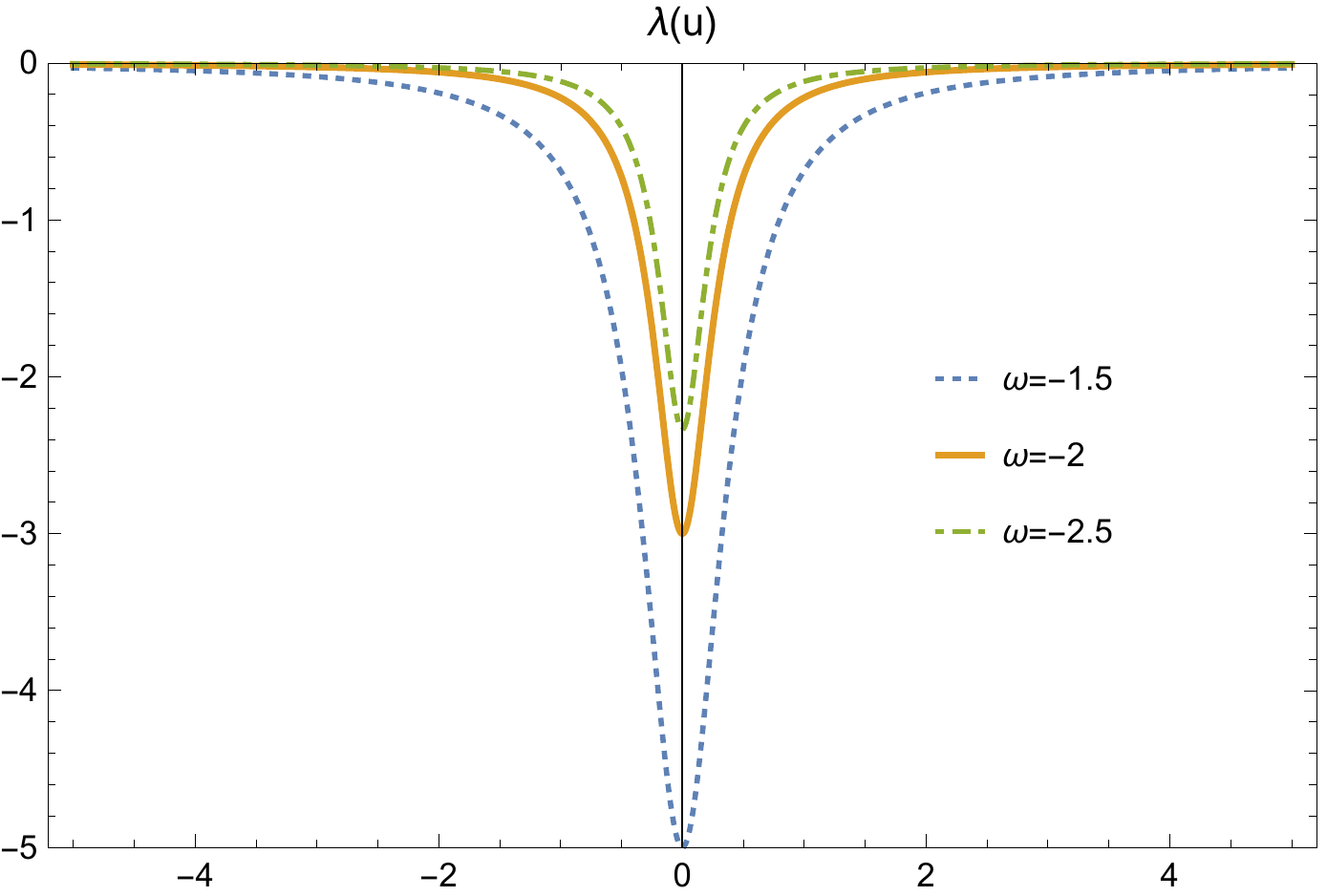}
     \includegraphics[width=0.47\linewidth]{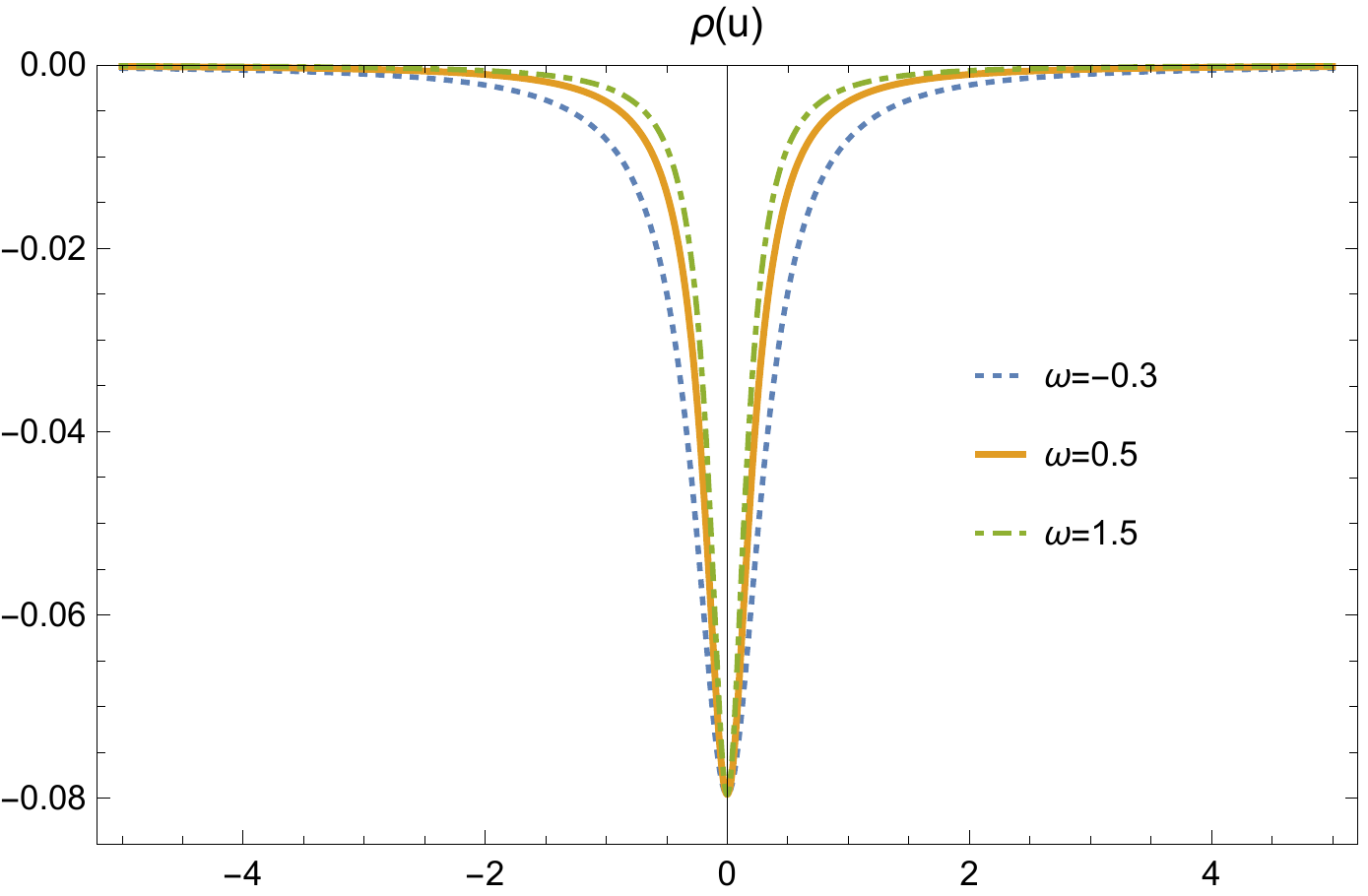}
    \includegraphics[width=0.47\linewidth]{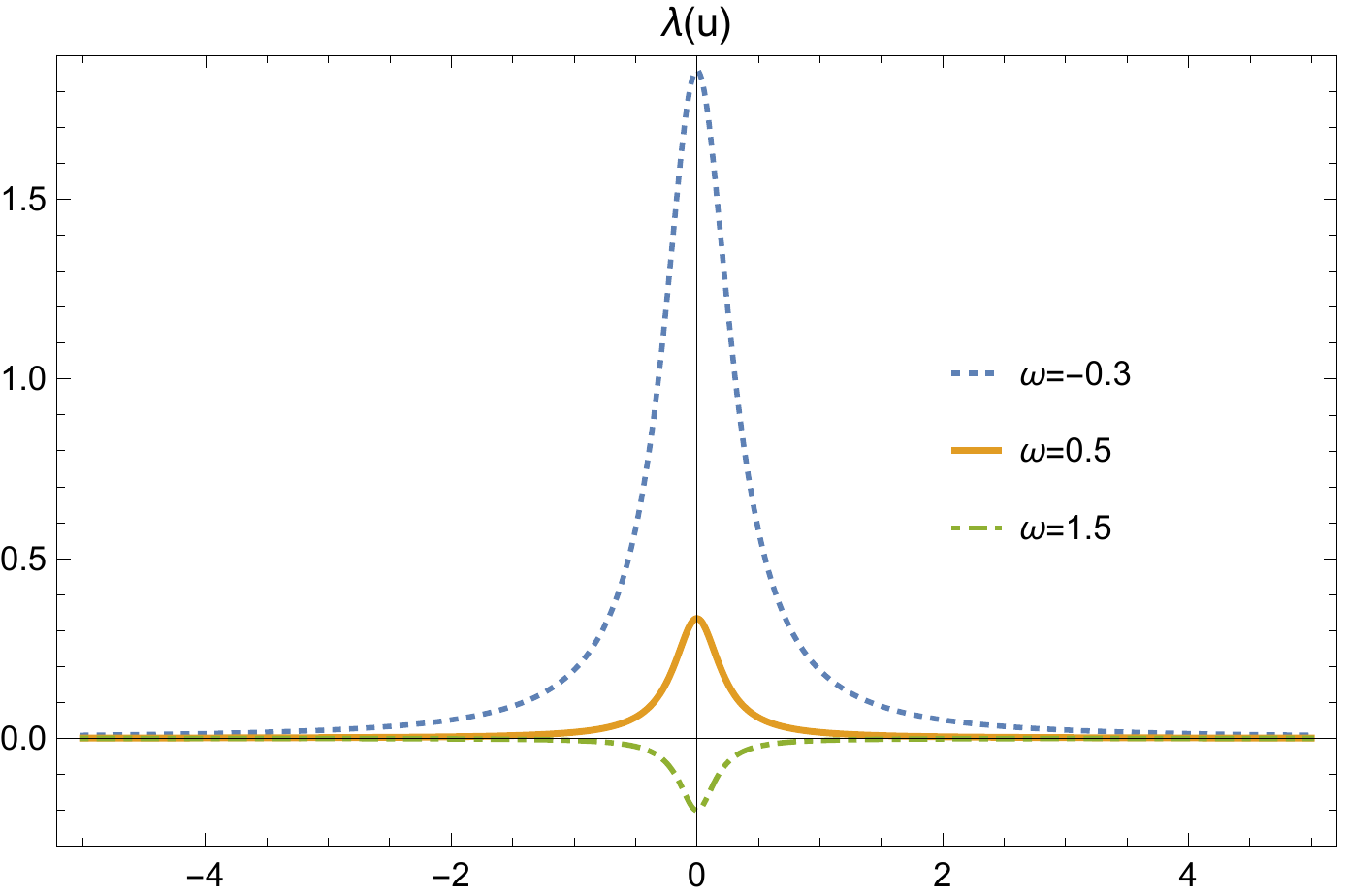}
    \caption{Results for the specific stress-energy profile given by \eqref{rhoSE} and \eqref{PrSE} with $a^2=-1/(4\pi \rho_0(1+\omega))$ and $\alpha=2$. The upper plots are for the case $\rho_0=1$ ($\omega<-1$), and the lower plots for $\rho_0=-1$ ($\omega>-1$).}
    \label{GraphSE}
\end{figure*}

We now consider the strategy of specifying the profile of the energy density and radial pressure given by:
\bea
\rho(u)=\rho_0 \left( \frac{a^2}{a^2 + u^2} \right)^\alpha \,,
\label{rhoSE}
	\\
p_r(u)=p_0 \left( \frac{a^2}{a^2 + u^2} \right)^\alpha \,,
\label{PrSE}
\eea
with $\alpha > 0$, so that both components tend to zero at spatial infinity. In addition to this, we close the system by considering the specific choice for the metric function
\be
R(u)=\sqrt{a^2+u^2}\,.
\label{choiceR}
\ee
Note that Eqs. (\ref{rhoSE}) and (\ref{PrSE}) can be written as $p_r=\omega \rho$, with $\omega = p_0/\rho_0$. Thus, this case is formally equivalent to choosing Eq. (\ref{choiceR}), one of Eqs. (\ref{rhoSE}) or (\ref{PrSE}), and the equation of state $p_r(u)=\omega \rho(u)$.

The gravitational field equations \eqref{rhou}-\eqref{ptu} provide the following solutions:
\begin{eqnarray}
A(u)&=&-4 \pi  a^2 \text{$\rho $}_0 (\omega +1) \left(\frac{a^2}{a^2+u^2}\right)^{\alpha -2 } \,, 
   	\\
\lambda(u)&=& 1+ 4 \pi  \text{$\rho $}_0  \left[2 a^2 
\omega -(2\alpha -5) u^2 (\omega +1)\right] \times
	\nonumber \\
&& \qquad \times 
\left(\frac{a^2}{a^2+u^2}\right)^{\alpha -1} \,.
\end{eqnarray}
As before, we impose the asymptotic flatness condition, namely,
\bea
\lim_{u\rightarrow\pm\infty}A(u)\rightarrow 1  \,,
\eea
and taking into account that $A(u)$ should be positive and regular $\forall u$, implies the following two stringent restrictions:
\bea
\alpha=2, \qquad  \text{and}  \qquad 4 \pi  a^2 \rho_0(1+\omega)=-1,
\eea
where the second condition imposes:
\bea
\rho_0(1+\omega)<0.
\label{condition}
\eea
This implies two specific cases: (i) $\rho_0 > 0$ and $\omega<-1$, so that taking into account the equation of state $\omega = p_0/\rho_0$, implies a negative radial pressure at the throat; or (ii) $\rho_0 < 0$ and $\omega > -1$, so that $p_0 > 0$ for $-1 < \omega <0$, and $p_0 < 0$ for $\omega >0$. Specific cases are depicted in Fig. \ref{GraphSE}.
Relative to the analysis at the throat, note that the wormhole conditions are satisfied, namely, $R'(u_0)=0$ and $R''(0)=1/|a|>0$. In addition to this, for the imposition of the asymptotic flatness condition, namely, $\alpha=2$, we readily obtain $A(u)=1$, so that $A'(u_0)=A''(u_0)=0$.

\subsection{Black bounce solutions}

Recently, a number of novel regular ``black-bounce'' spacetimes were explored \cite{Simpson:2018tsi,Lobo:2020ffi}. These are specific geometries where the ``area radius'' always remains non-zero, thereby leading to a ``throat'' that is either timelike (corresponding to a traversable wormhole), spacelike (corresponding to a ``bounce'' into a future universe), or null (corresponding to a ``one-way wormhole''). The regularity, the energy conditions, and the causal structure of these models were analysed in detail in Refs. \cite{Simpson:2018tsi,Lobo:2020ffi}. The main results are several new geometries with two or more horizons, with the possibility of an extremal case. Motivated by these novel solutions, in this subsection we shall analyse specific generalized ``black-bounce'' wormhole geometries induced by action-dependent Lagrangian theories.

\subsubsection{Simpson--Visser black-bounce spacetime}

\begin{figure*}[htbp!]
    \centering
    \includegraphics[width=0.45\linewidth]{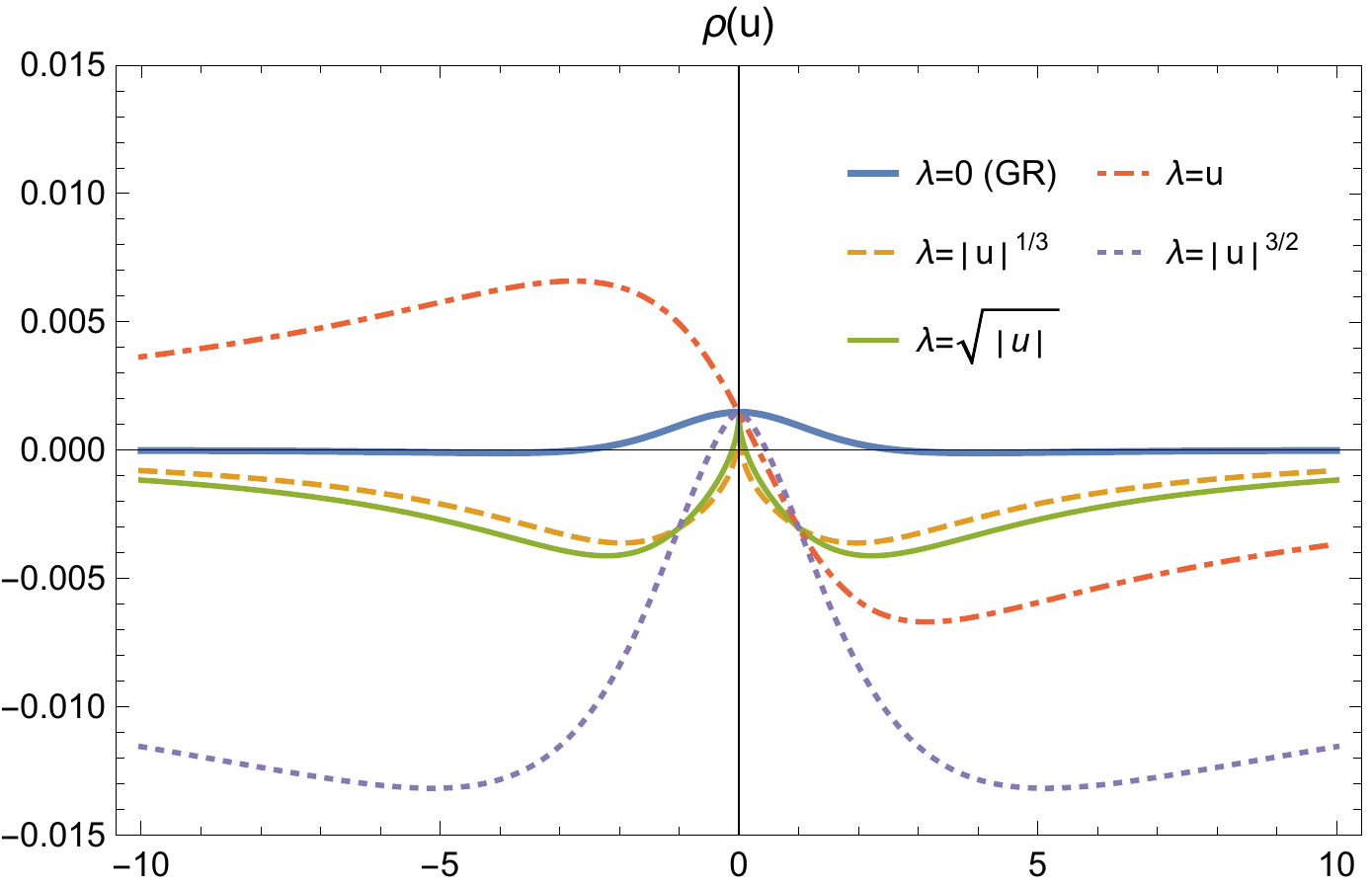}
    \includegraphics[width=0.45\linewidth]{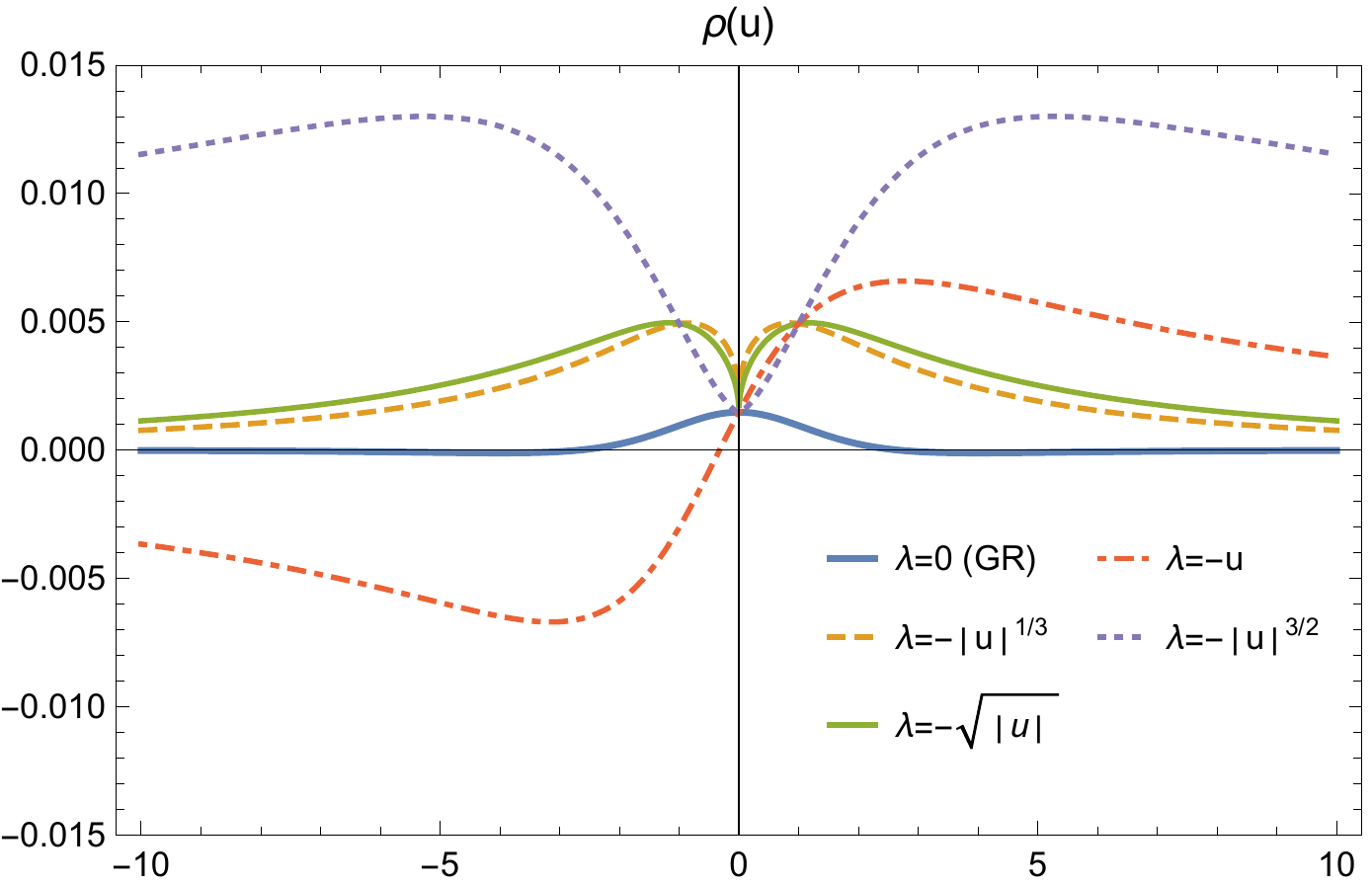}
    \includegraphics[width=0.45\linewidth]{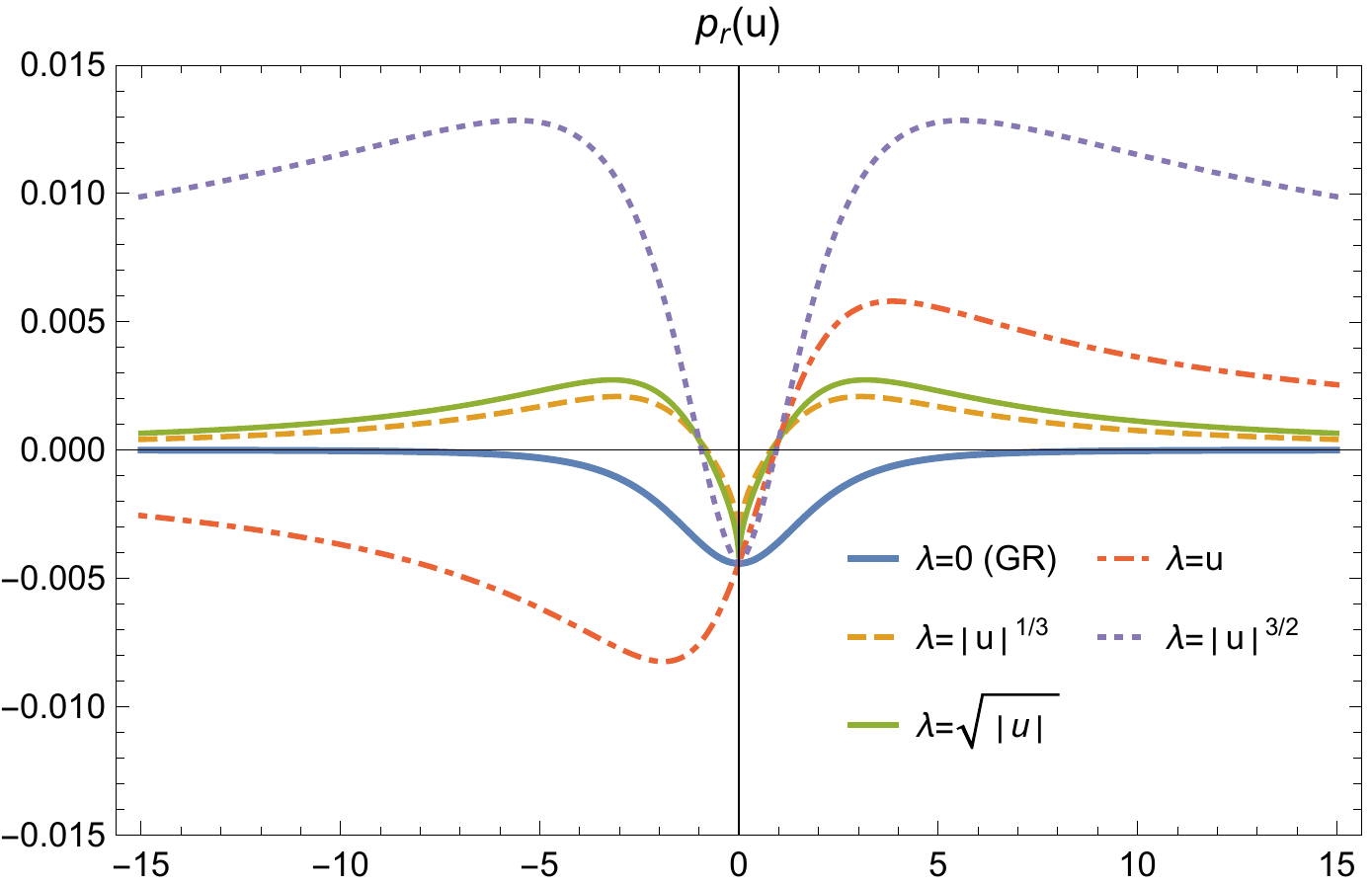}
    \includegraphics[width=0.45\linewidth]{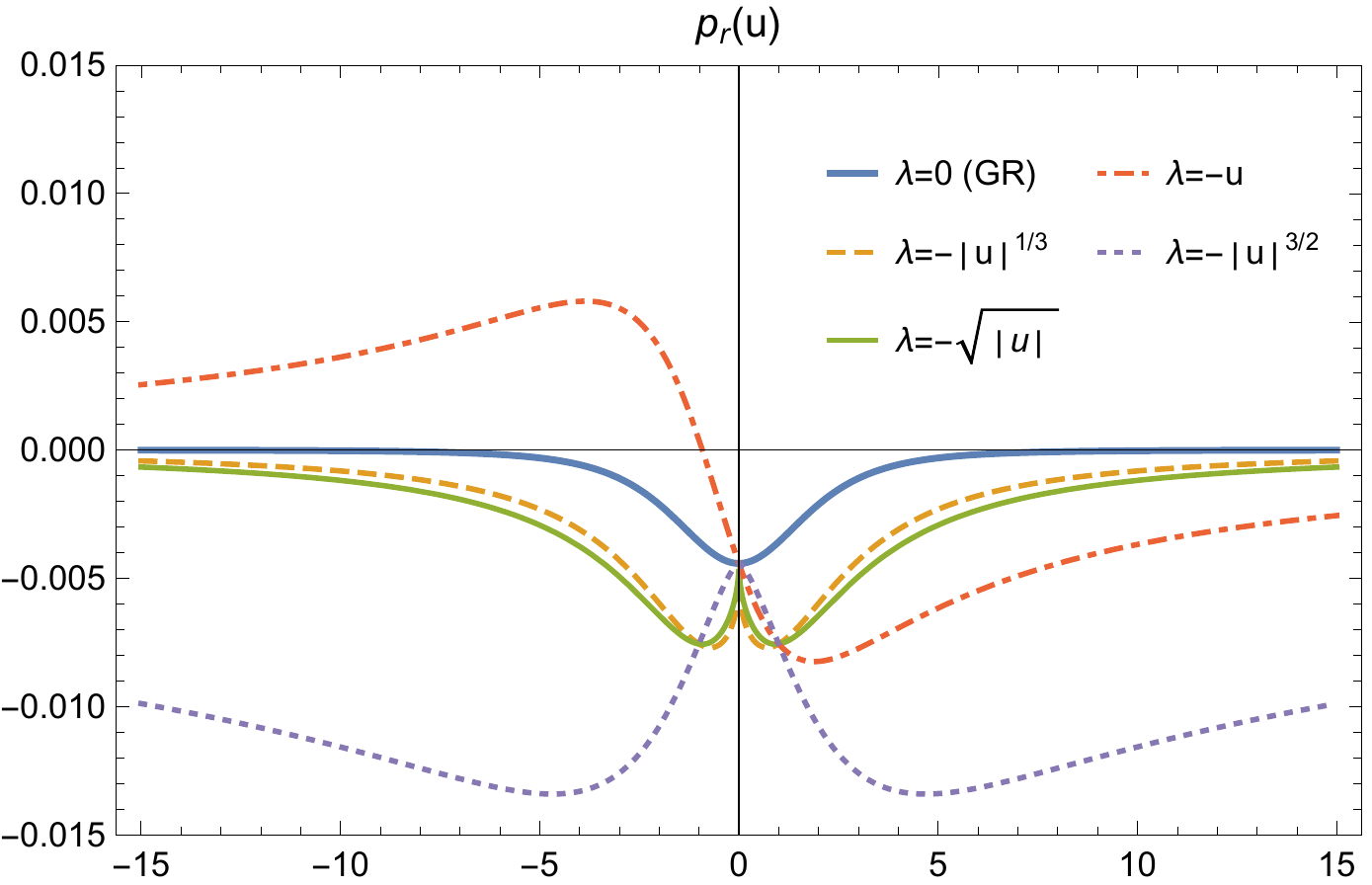}
    \caption{The plots depict the Simpson-Visser black bounce solution, for the specific choices 
    $m=1, a=3 $. Note that one may obtain wormhole configurations with an entirely positive energy density throughout the spacetime, for negative values of the function $\lambda(u)$; and positive radial pressures in the negative branch of the $u$-axis. As before, it is transparent from the plots that these compact objects possess a richer geometrical structure than their general relativistic counterparts. See the text fro more details.}
    \label{GraphSV}
\end{figure*}

In this section, we consider a specific black bounce geometry, which we denote as the Simpson-Visser solution \cite{Lobo:2020ffi,Lobo:2020kxn,Simpson:2018tsi}. 
Consider the following parameters, which were presented in Ref. \cite{Simpson:2018tsi}:
\begin{eqnarray}
R(u)=\sqrt{u^2+a^2},\qquad A(u)=1-\frac{2m}{\sqrt{u^2+a^2}}\,.\label{Visser}
\end{eqnarray}
Note that the Schwarzschild solution is recovered, if we take the limit $a\rightarrow 0$.

This spacetime possesses several interesting properties \cite{Simpson:2018tsi}. First, for $a>0$ the geometry is everywhere regular, which can verified as $R(u)$ is never zero, and is regular, as is $A(u)$. Now, one has several cases, for instance: 
(i) if $0<a<2m$, two horizons exist, namely, $u_{\pm}=\pm\sqrt{(2m)^2-a^2}$, where $u_{+}$ is positive and $u_{-}$ is negative. This solution corresponds to a regular black hole spacetime, where the core consists of a bounce located at $u=0$; 
(ii) if $a=2m$, a wormhole exists with a throat located at $u=0$. This is an extremal null throat, which can only be crossed from one region to another, so that the wormhole is only one-way traversable; 
(iii) finally, if $a>2m$, a two-way traversable wormhole exists, that possesses a timelike throat located at $u=0$. Thus, only the case (c) for $a>2m$ interests us here. We refer the reader to Refs. \cite{Simpson:2018tsi,Lobo:2020ffi} for specific details.

Taking into account the choices for the metric functions (\ref{Visser}), the field equations \eqref{rhou}-\eqref{ptu} provide the following stress-energy profile:
\be
\rho(u)_{SV}=- \frac{1}{8 \pi } \left[\frac{a^2 \left(\sqrt{a^2+u^2}-4 m\right)}{\left(a^2+u^2\right)^{5/2}}+\frac{\lambda
   (u)}{a^2+u^2} \right] \,,
\ee   
\be   
   p_r(u)_{SV}=\frac{\left(a^2+u^2\right) \lambda (u)-a^2}{8 \pi \left(a^2+u^2\right)^2} \,,
\ee
\be   
   p_t(u)_{SV}=\frac{a^2 \left(\sqrt{a^2+u^2}-m\right)}{8 \pi \left(a^2+u^2\right)^{5/2}} \,,
\ee
respectively. The asymptotic limits of the energy density and the radial pressure are given by:
\be
\lim_{u\rightarrow \pm\infty}\rho(u)_{SV} \sim - \lim_{u\rightarrow \pm\infty}p_r(u)_{SV}  \sim - \lim_{u\rightarrow \pm\infty}\frac{\lambda(u)}{a^2+u^2} .
\ee
As before, if we assume a power law for $\lambda(u) \sim u^\alpha$, the asymptotic flatness condition and the regularity of the stress-energy components, as before, imposes that $0 \leq \alpha <2$; note that the tangential pressure $p_t(u)\rightarrow 0$ for $u\rightarrow \pm \infty$, and possesses a maximum at the wormhole throat, i.e., $p_t(u=0)= \left(a-m\right)/(8 \pi a^3) $, and is positive throughout the spacetime as we are only considering the condition $a>2m$.

Several choices for the function are depicted in Fig. \ref{GraphSV}. Depending on the sign of $\lambda(u)$, one obtains a plethora of specific symmetric or asymmetric solutions. Note that
these compact objects possess a richer geometrical structure than their general relativistic counterparts. It is transparent from Fig. \ref{GraphSV} that one may obtain wormhole configurations with an entirely positive energy density throughout the spacetime, for negative values of the function $\lambda(u)$; for instance, for the latter it is also possible to obtain positive radial pressures in the negative branch of the $u$-axis. However, the NEC is always violated at the wormhole throat.

\subsubsection{Black bounce II}\label{sec:bbII}

\begin{figure*}[htbp!]
    \centering
    \includegraphics[width=0.45\linewidth]{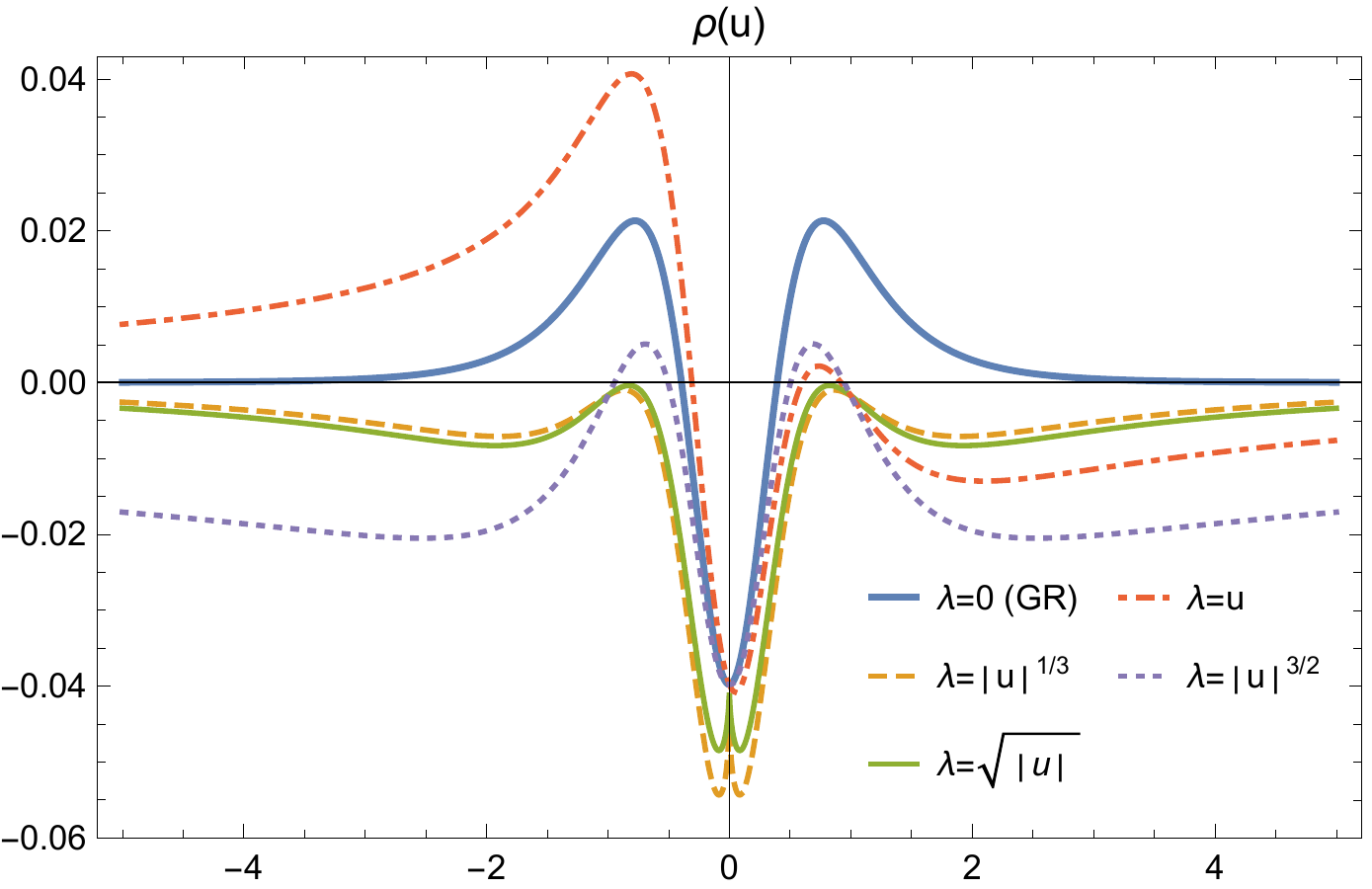}
    \includegraphics[width=0.45\linewidth]{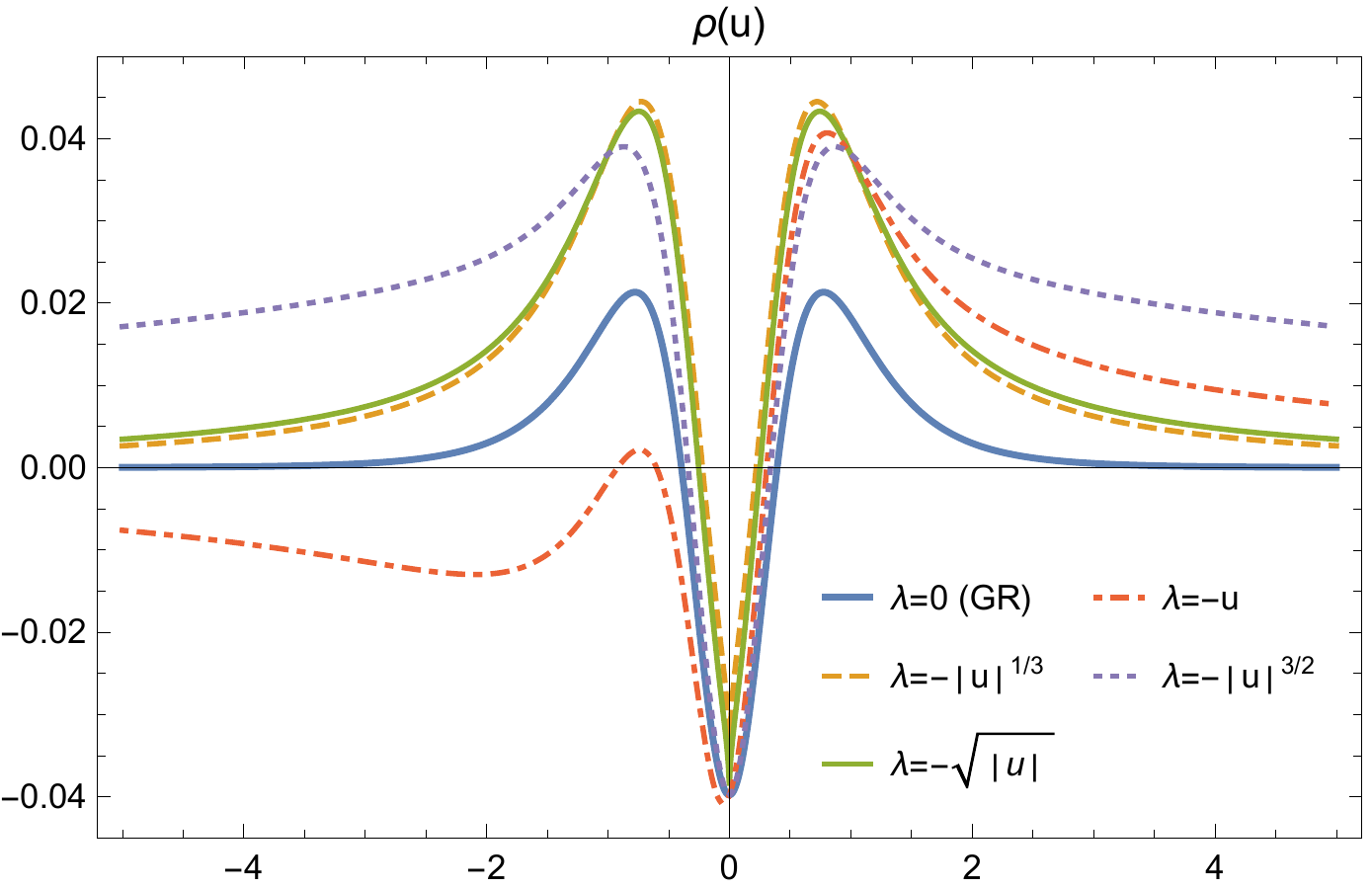}
    \includegraphics[width=0.45\linewidth]{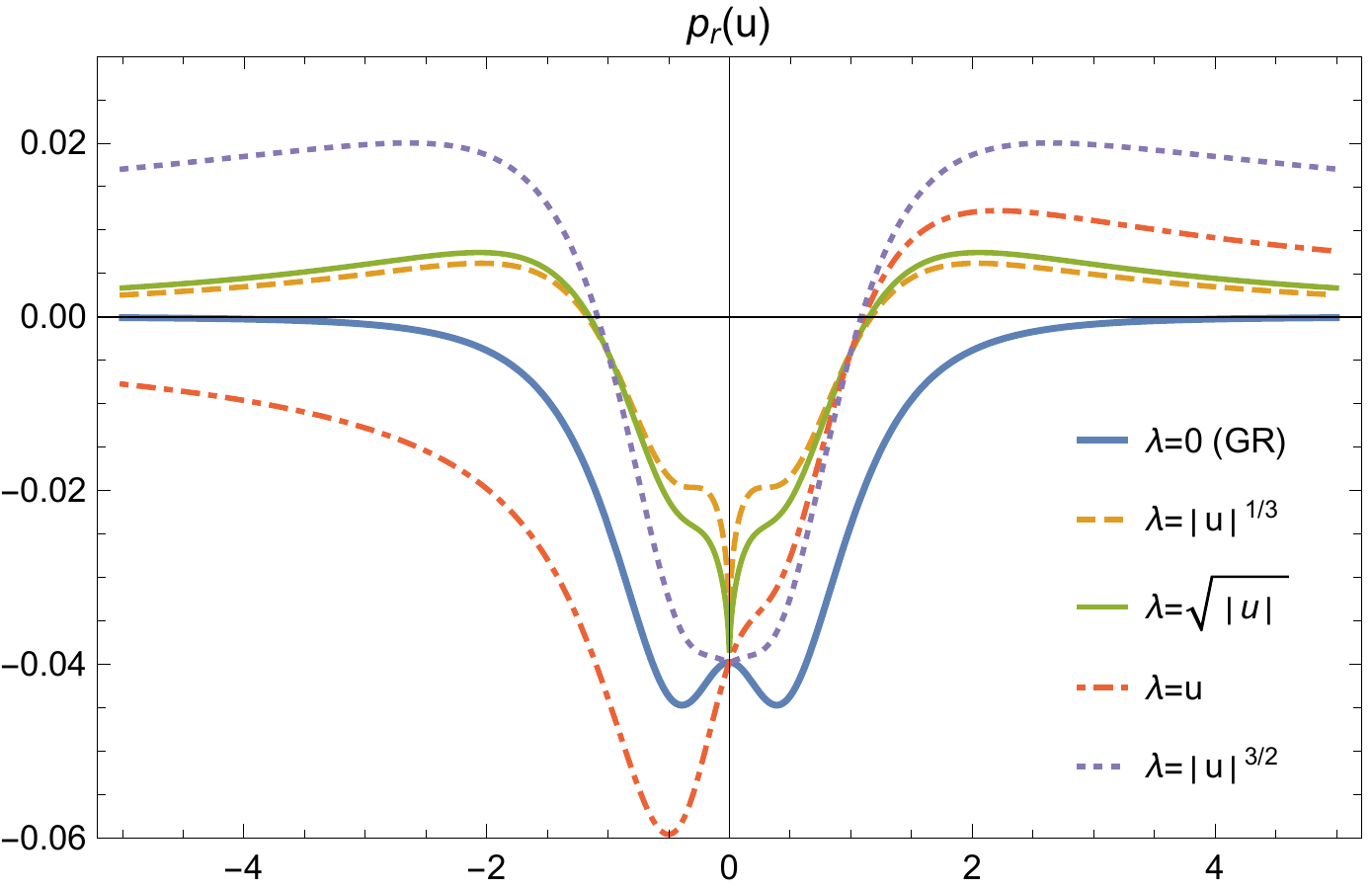}
    \includegraphics[width=0.45\linewidth]{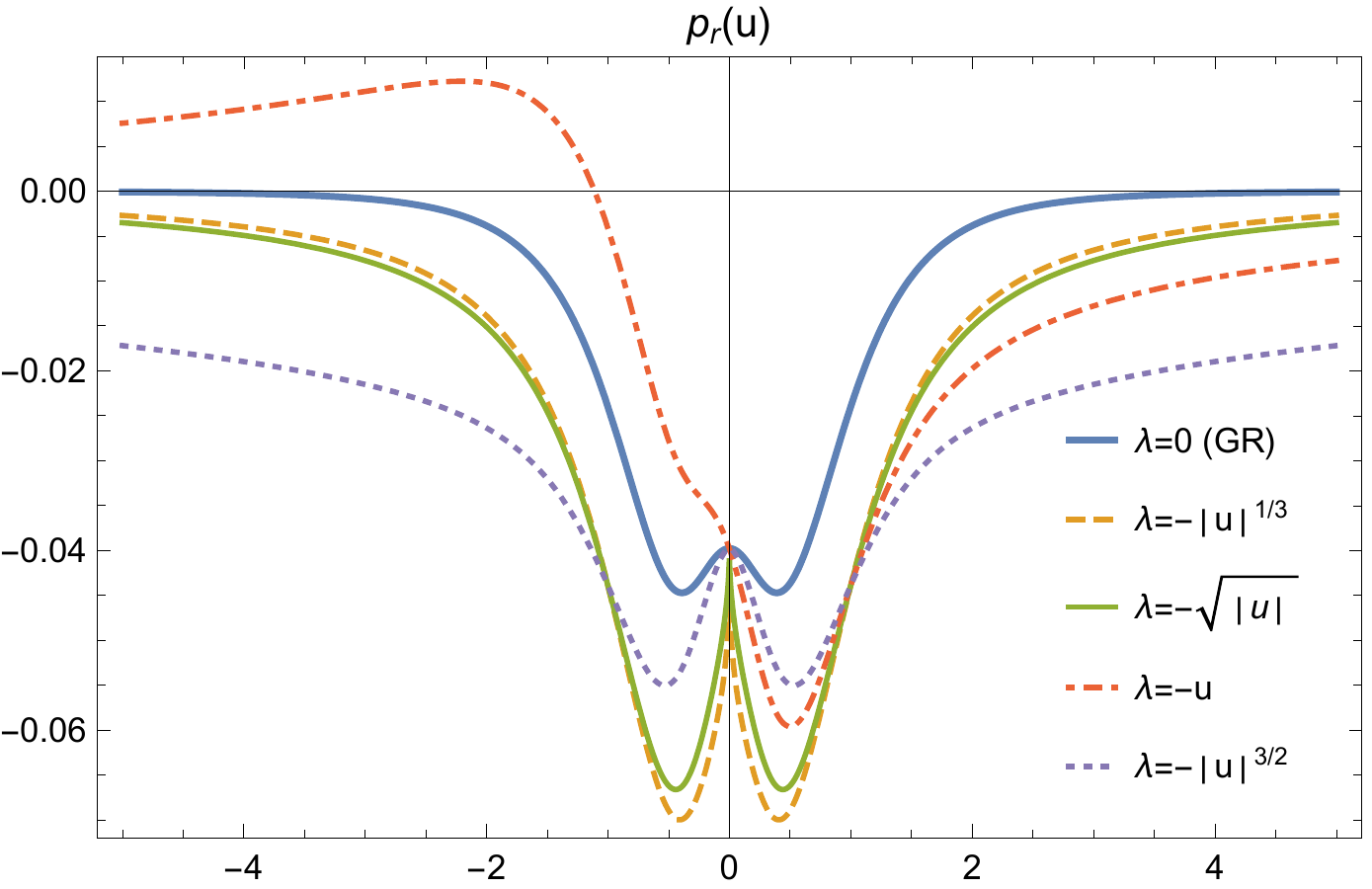}
    \includegraphics[width=0.45\linewidth]{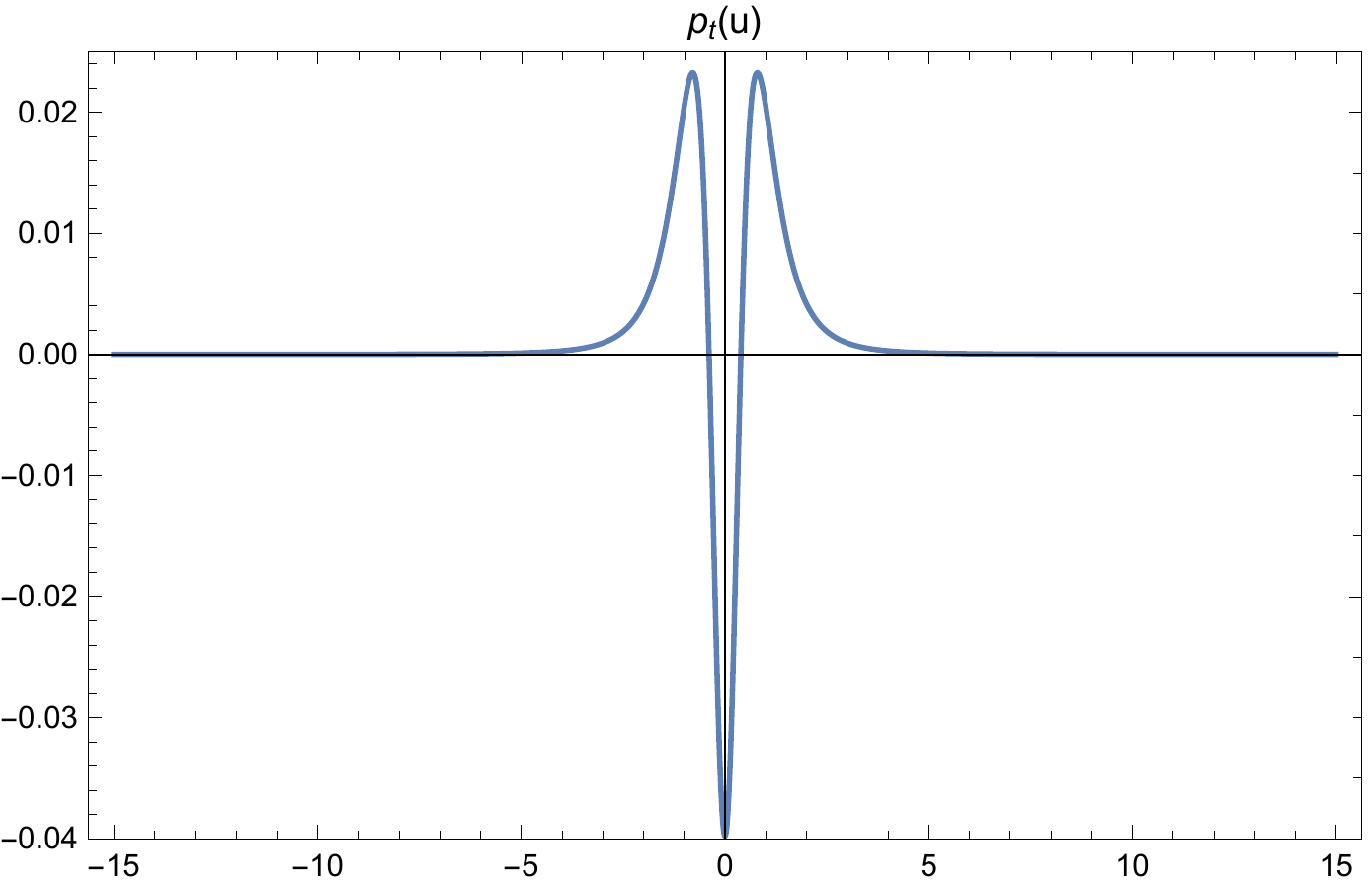}
    \caption{The plots depict the stress-energy profile for the black bounce II solution, given by the metric function (\ref{mod2}), where for numerical convenience we have assumed the following choices for the parameters: $m=1, a=1$. Recall that the parameters are restricted by the condition $a>4m/(3\sqrt{3})$. As before these wormhole configurations possess a far richer internal structure than their general relativistic counterparts, and depending on the sign of $\lambda(u)$, specific symmetric or asymmetric solutions are obtained. However, here the tangential pressure at the wormhole throat, $p_t(u=0)= \left(a-2m\right)/(8 \pi a^3) $, takes negative values for the parameter range $4m/(3\sqrt{3})< a <2m$, and positive values for $a > 2m$. We refer the reader to the text for more details.}
    \label{GraphBBII}
\end{figure*}

Another black bounce spacetime that exhibits interesting properties, is given by the following specific metric functions \cite{Lobo:2020ffi}:
\be
R(u)=\sqrt{u^2+a^2},\qquad A(u)=1-\frac{2mu^2}{(u^2+a^2)^{3/2}}\label{mod2}\,.
\ee
Note that by solving for the roots of the function $A(u)=0$, we have: i) for $a<a_{\rm ext}=4m/(3\sqrt{3})$, there are four real solutions, which are symmetrical to each other, namely, $(r_+,r_C,-r_C,-r_+)$, (where $u_+$ corresponds to the event horizon and $u_C$ to a Cauchy horizon); ii) for $a=a_{\rm ext}$, we have two real solutions $(u_+,-u_+)$; iii) and for $a>a_{\rm ext}$, no real value exists. We refer the reader to \cite{Lobo:2020ffi} for more details. Thus, in order to have a traversable wormhole solution, where only the case of $A(u)>0$ is satisfied, only the specific case for $a>a_{\rm ext}$ interests us here.

For this case, the gravitational field equations \eqref{rhou}-\eqref{ptu} yield the stress-energy profile, given by the following relations:
\be
\rho(u)=- \frac{1}{8 \pi } \left[ \frac{a^2 \left(a^2+u^2\right)^{3/2} - 8 m a^2 u^2}{\left(a^2+u^2\right)^{7/2}}+\frac{\lambda (u)}{a^2+u^2} \right],
\ee
\be   
   p_r(u)= \frac{1}{8 \pi } \left[ - \frac{a^2 \left(a^2+u^2\right)^{3/2}+4 m a^2 u^2}{ \left(a^2+u^2\right)^{7/2}}
   +\frac{\lambda (u)}{a^2+u^2} \right],
\ee
\be   
   p_t(u)=\frac{a^2 u^2 \left(\sqrt{a^2+u^2}+5 m\right)+a^4 \left(\sqrt{a^2+u^2}-2 m\right)}{8 \pi 
   \left(a^2+u^2\right)^{7/2}},
\ee
respectively, which are depicted in Fig. \ref{GraphBBII} for specific choices of the model parameters.

Assume, once again, a power law for $\lambda(u) \sim u^\alpha$, we have that $0 \leq \alpha <2$ by the asymptotic flatness condition and the regularity of the stress-energy components, as before. 
As before these wormhole geometries induced by action-dependent Lagrangian theories possesses a far richer internal structure than their general relativistic counterparts, and depending on the sign of $\lambda(u)$, specific symmetric or asymmetric solutions are obtained. We refer the reader to 
Fig. \ref{GraphBBII} for a qualitative behaviour of the stress-energy profile; recall that taking into account the parameter range, we consider the condition $a>4m/(3\sqrt{3})$. Here the tangential pressure at the wormhole throat is given by $p_t(u=0)= \left(a-2m\right)/(8 \pi a^3) $, and takes negative values for  $4m/(3\sqrt{3})< a <2m$, possessing a minimum at the throat, and positive values for $a > 2m$; note that $p_t(u)\rightarrow 0$ for $u\rightarrow \pm \infty$.

\subsubsection{Black bounce III}\label{sec:bbIII}

\begin{figure*}[htbp!]
    \centering
    \includegraphics[width=0.45\linewidth]{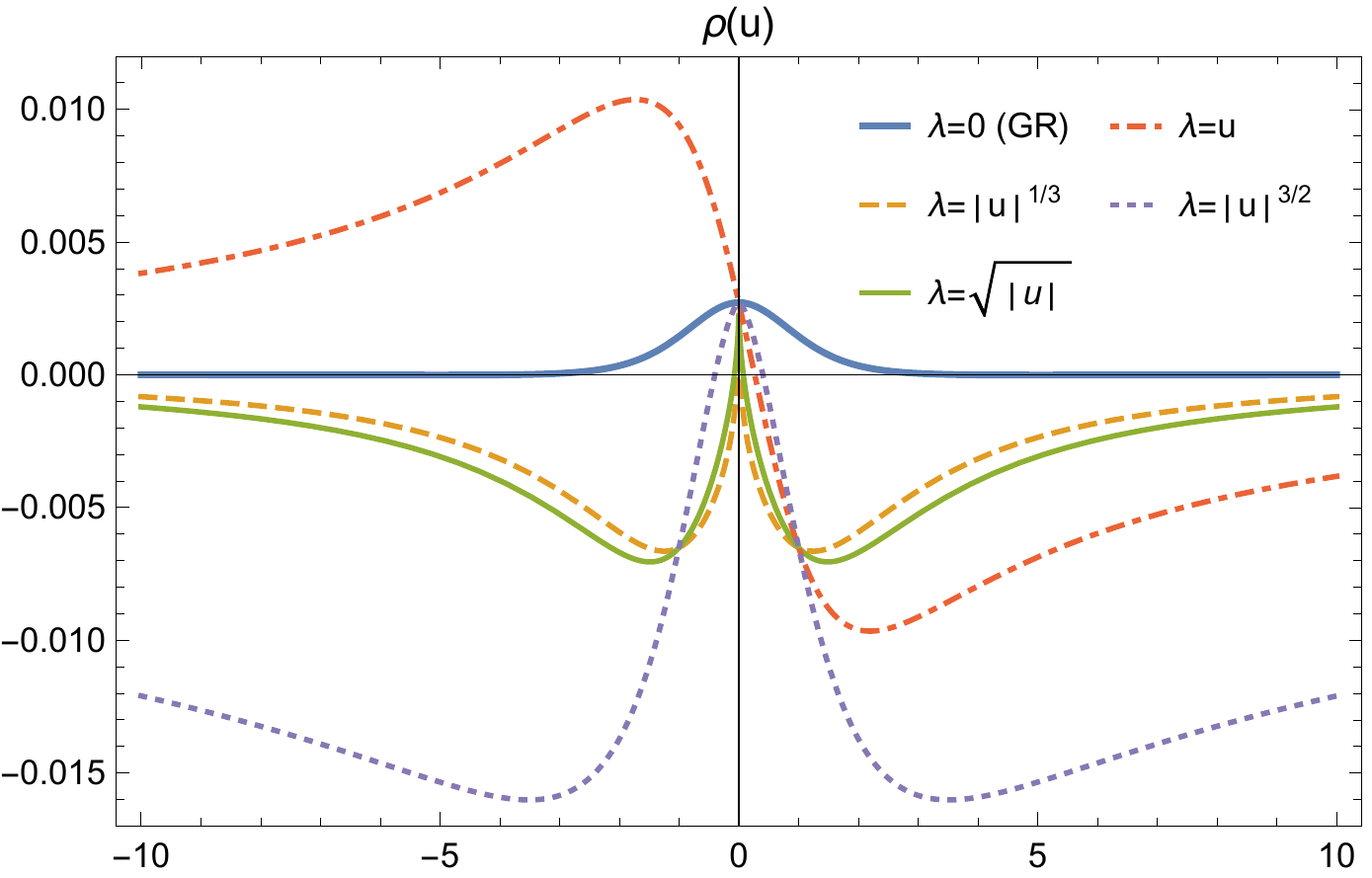}
    \includegraphics[width=0.45\linewidth]{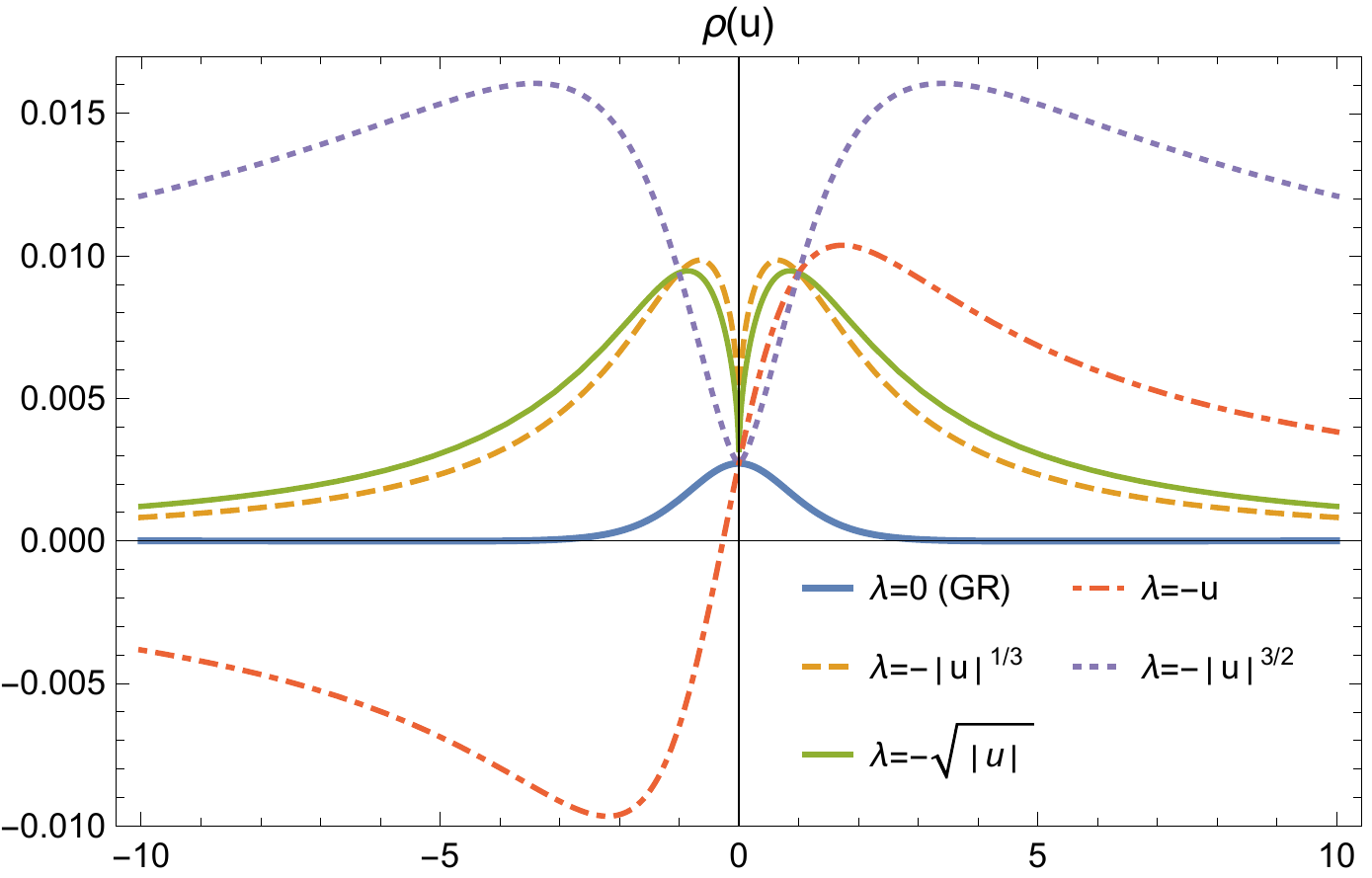}
    \includegraphics[width=0.45\linewidth]{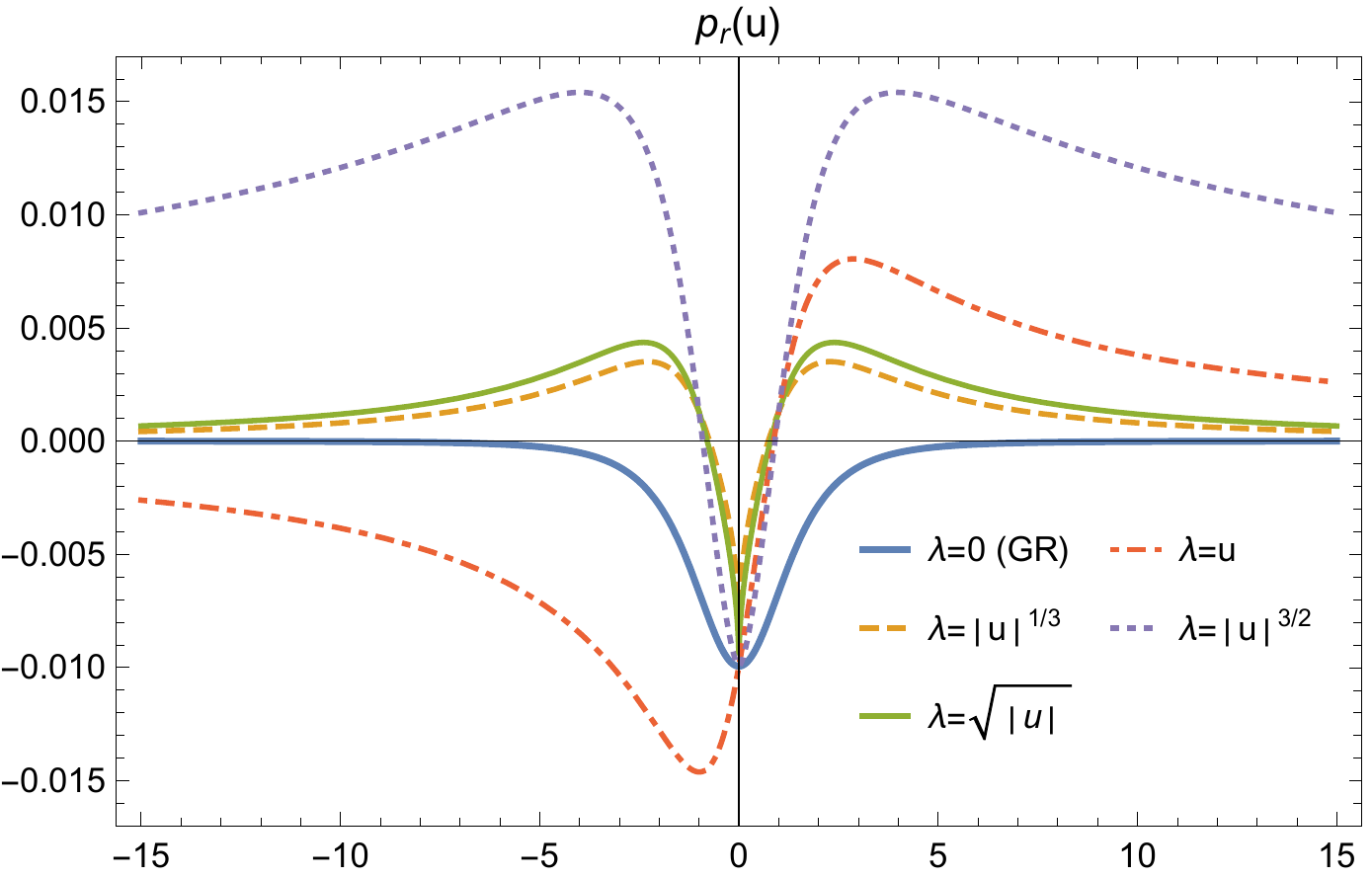}
    \includegraphics[width=0.45\linewidth]{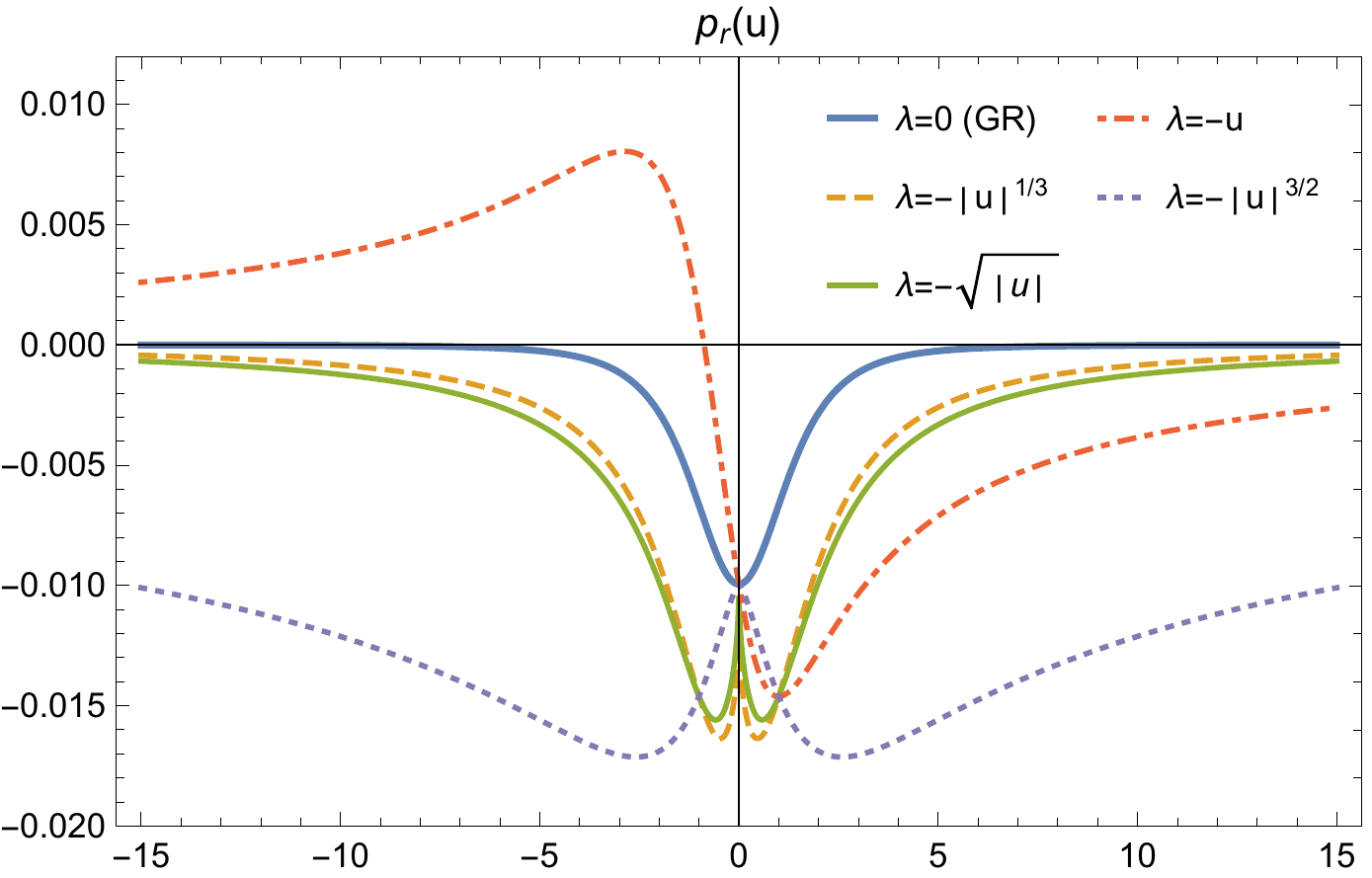}
    \includegraphics[width=0.45\linewidth]{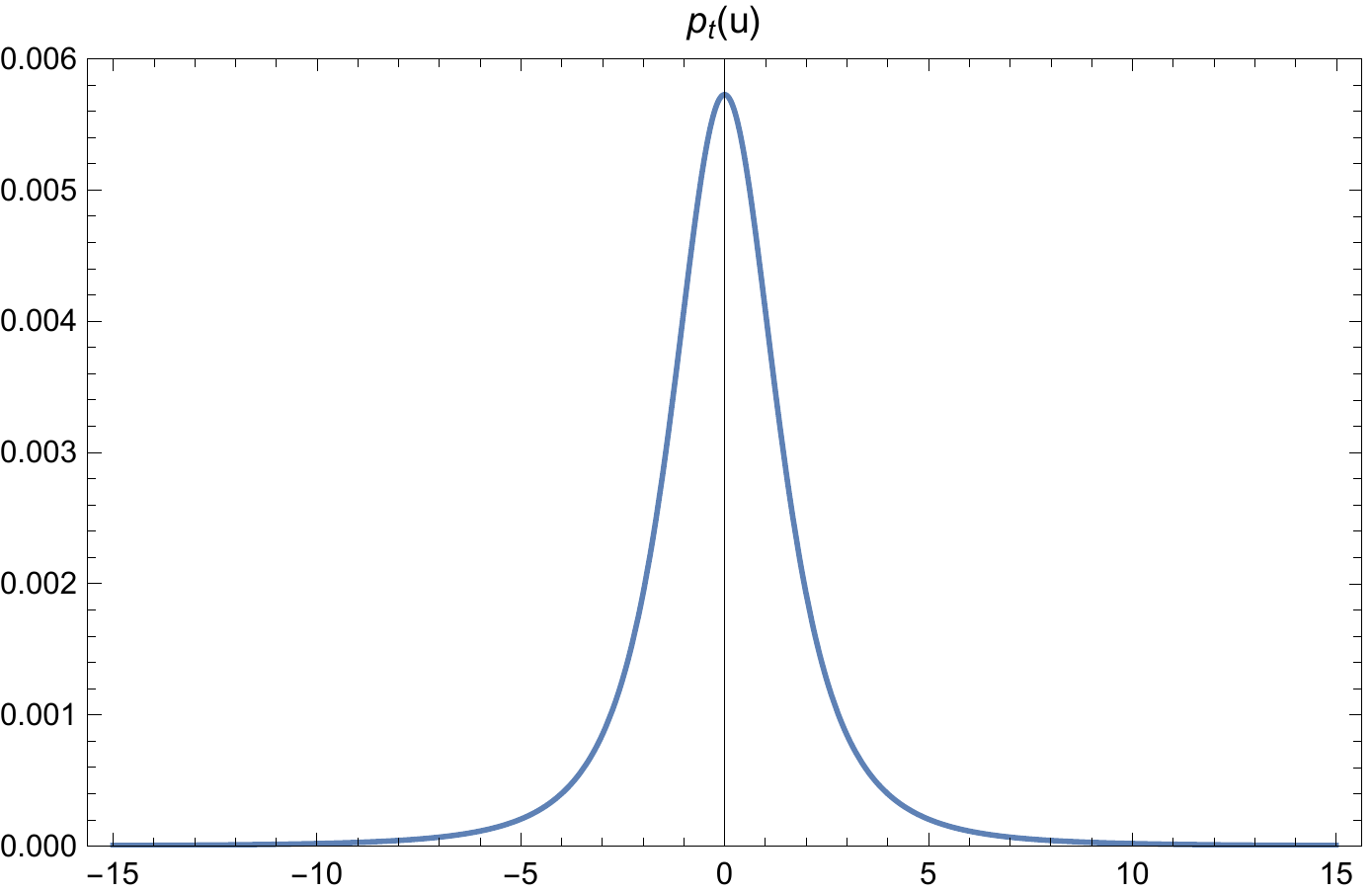}
    \caption{The plots depict the stress-energy profile for the black bounce III solution, given by the functions (\ref{mod4a}) and (\ref{mod4}), for the case $n=1$, where $a>a_{\rm ext}$, so that there are no event horizons. We have chosen the following values for the parameters: $m=1$ and $a=2$. Note that for these solutions, the negative energy densities are improved, and one may also obtain positive radial pressures for positive values of the function $\lambda(u)$. See the text for more details.}
    \label{GraphBBIII}
\end{figure*}

Finally, we consider another black bounce solution explored in \cite{Lobo:2020ffi}, that also exhibits interesting properties, given by
\be
R(u)=\sqrt{u^2+a^2}\,.
\label{mod4a}
\ee
and the mass function:
\begin{equation}
    M(u)=m \left(\frac{R(u)}{u}\right)\left(\frac{2}{\pi}\right)^n \arctan^n\left(\frac{u}{a}\right) \,,
\end{equation}
so that the metric function $A(u)$ is given by
\begin{eqnarray}
A(u)&=&1-\frac{2M(u)}{R(u)}
    \nonumber \\
&=&1- \frac{2m}{u} \left(\frac{2}{\pi}\right)^n \arctan^n\left(\frac{u}{a}\right)
\label{mod4}\,.
\end{eqnarray}
In the limit $(a,n)\rightarrow 0$ we regain the Schwarzschild solution. However, one can fix $n$ and regulate the presence of horizons by adjusting $a$. For instance, consider $n=1$, where the extreme case is given by $a_{\rm ext}=4m/\pi$ \cite{Lobo:2020ffi}.

The causal structure, for the specific case of $n=1$, is given by: (i) for $a>a_{\rm ext}$, this  corresponds to the traditional two-way traversable wormhole; (ii) for $a=a_{\rm ext}$, we have a one-way wormhole geometry with an extremal null throat; (iii) for $a<a_{\rm ext}$, we have one horizon located in each universe, where one may propagate through this event horizon, located at $u=u_+$ in order to reach the spacelike ``bounce'' hypersurface at $u=0$, before ``bouncing'' into a future version of our own universe. We refer the interested reader to Ref. \cite{Lobo:2020ffi} for more details. Thus, we are only interested in the case $a>a_{\rm ext}$, for $n=1$, where there are no event horizons.

Taking into account the metric functions (\ref{mod4a}) and (\ref{mod4}), the gravitational field equations \eqref{rhou}-\eqref{ptu} provide the following stress-energy components:
\begin{widetext}
\begin{eqnarray}
\rho(u)&=&-\frac{1}{8\pi}\left\{ \frac{a^2}{\left(a^2+u^2\right)^2}\left[1-\frac{2m}{ua}  
\left(\frac{2}{\pi}\right)^n
 \left(\arctan\left(\frac{u}{a}\right)\right)^{n-1} \left(a \arctan\left(\frac{u}{a}\right)+n
   u\right)\right]+\frac{ \lambda (u)}{a^2+u^2}\right\},
\end{eqnarray}
\begin{eqnarray}
   p_r(u)&=&\frac{1}{8\pi}\left\{ \frac{2ma}{u \left(a^2+u^2\right)^2} 
		\left(\frac{2}{\pi}\right)^n		
		\left(\arctan\left(\frac{u}{a}\right)\right)^{n-1}
   \left[a \arctan\left(\frac{u}{a}\right)-n u\right] +\frac{ \left(a^2+u^2\right) \lambda (u)-a^2}{(a^2+u^2)^2}\right\},
\end{eqnarray}
\begin{eqnarray}
   p_t(u)&=&\frac{a^2}{8\pi (a^2+u^2)^2}\left\{1-  
		\frac{m}{au^3}\left(\frac{2}{\pi}\right)^n   
   \left(\arctan\left(\frac{u}{a}\right)\right)^{n-2}\right.\times	\nonumber \\
   &&   
   \qquad \times
   \left.\left[2 \left(a^3+2 a u^2\right) \left(\arctan\left(\frac{u}{a}\right)\right)^2-2 n u
   \left(a^2+u^2\right) \arctan\left(\frac{u}{a}\right)+a (n-1) n u^2\right]\right\}.
\end{eqnarray}
\end{widetext}
respectively. Here, we will only consider, for simplicity, the specific case of $n=1$ and $a>a_{\rm ext}$.

If we consider, as before, a power law for $\lambda(u) \sim u^\alpha$, the asymptotic flatness and regularity conditions impose $0 \leq \alpha <2$. Once again, one obtains a wide variety of solutions, both symmetric and asymmetric, which are depicted in Fig. \ref{GraphBBIII}. The advantage of these solutions consists essentially in that they ameliorate the negative energy densities for negative values of the function $\lambda(u)$. However, positive values of the function $\lambda(u)$ allow positive radial pressures as is transparent in Fig. \ref{GraphBBIII}.
The tangential pressure tends to zero at spatial infinity, i.e., $p_t(u)\rightarrow 0$ for $u\rightarrow \pm \infty$, and possesses a maximum at the throat, as depicted in Fig. \ref{GraphBBIII}.

\section{Conclusions}\label{sec:conclusions}

In this work, we have explored wormhole geometries in the recently proposed action-dependent Lagrangian theories \cite{Lazo:2017udy}, that are obtained through an action principle for action-dependent Lagrangians by generalizing the Herglotz variational problem for several independent variables. An interesting feature of these theories as compared with previous implementations of dissipative effects in gravity is the possible arising of such phenomena from a least action principle, so they are of a purely geometric nature. It was shown that the generalized gravitational field equation essentially depends on a background four-vector $\lambda^\mu$, that plays the role of a coupling parameter associated with the dependence of the gravitational Lagrangian upon the action, and may generically depend on the spacetime coordinates. 
In the context of wormhole configurations, we have used the ``Buchdahl coordinates'', and found that the four-vector is given generically by $\lambda_{\mu}=\left(0,0,\lambda_{\theta}(u,\theta),0\right)$. In addition to this restriction the spacetime geometry is also severely constrained by the condition $g_{tt}g_{uu}=-1$, where $u$ is the radial coordinate. 

More specifically, we have shown that the field equations (\ref{rhou})--(\ref{ptu}), impose a system of three independent equations with six unknown functions of the radial coordinate $u$, namely, $\rho(u)$, $p_r(u)$, $p_t(u)$, $A(u)$, $R(u)$ and $\lambda(u)$. Thus, one possesses several strategies to solve the system of equations.
For instance, one may consider a plausible stress-energy tensor profile by imposing equations of state $p_r=p_r(\rho)$ and $p_t=p_t(\rho)$, and close the system by adequately choosing the energy density, or a specific metric function. However, one may also adopt the reverse philosophy approach usually used in wormhole physics by simple choosing specific choices for the metric functions and $\lambda(u)$, and through the field equations determine the stress-energy profile responsible for sustaining the wormhole geometry. Here, we have found a plethora of specific asymptotically flat, symmetric and asymmetric, solutions with power law choices for the function $\lambda$, for instance, by generalizing the Ellis-Bronnikov solutions and the recently proposed black bounce geometries, amongst other solutions. We have shown that these compact objects possess a far richer geometrical structure than their general relativistic counterparts. It would be  interesting to investigate time-dependent spacetimes as outlined in \cite{KordZangeneh:2020jio,KordZangeneh:2020ixt} in order to explore the energy conditions. To this effect, one could consider that the metric functions in the line element (\ref{genmetric}) are now also time-dependent. This would imply a non-zero Einstein tensor $G^{t}_{u}$ component, which would consequently yield the presence of flux terms. Using the modified Einstein filed (\ref{fieldeq}), one could then expect that $Z^{t}_{u}\neq 0$, which would modify the structure of $\lambda_{\mu}$, implying a non-zero time-component, i.e., $\lambda_{t}\neq 0$. Work along these lines is presently underway.

\section*{Acknowledgments}
IA is funded by the Funda\c{c}\~ao para a Ci\^encia e a Tecnologia (FCT, Portugal) grant No. PD/BD/114435/2016 under the IDPASC PhD Program.
FSNL acknowledges support from the FCT Scientific Employment Stimulus contract with reference CEECIND/04057/2017. The authors also acknowledge funding from FCT Projects No. UID/FIS/04434/2020, No. CERN/FIS-PAR/0037/2019 and No. PTDC/FIS-OUT/29048/2017.




\end{document}